\documentclass[preprint,superscriptaddress,floatfix,amsmath,amssymb,amsfonts]{revtex4}

\usepackage{float}

\usepackage[normalem]{ulem}
\usepackage[english]{babel}
\usepackage{graphicx}
\usepackage{dcolumn}
\usepackage{bm}
\usepackage{blindtext}
\usepackage{slashed}
\usepackage{verbatim}
\usepackage{mathrsfs}
\usepackage{musicography}
\usepackage{amsmath}
\usepackage{cancel}
\usepackage{physics}
\usepackage{epstopdf}
\usepackage{mathtools}
\usepackage{tensor}
\usepackage{color}
\usepackage[usenames,dvipsnames]{pstricks}
\usepackage{epsfig}
\usepackage{pst-grad}
\usepackage{pst-plot}
\usepackage{hyperref}
\usepackage{verbatim}
\usepackage{tikz-feynman}
\usepackage{braket}
\usepackage{soul}
\usepackage{tensor}

\usetikzlibrary{decorations.markings}

\newcommand{\ii}{i}
\newcommand{\M}{\mathcal{M}}

\begin{document}

\title{A path integral formulation for particle detectors: \\the Unruh-DeWitt model as a line defect}

\author{Ivan M. Burbano\footnote{Now at the Department of Physics, University of California, 366 Physics North MC 7300, Berkeley,
CA 94720-7300, U.S.A.}}
\email{ivan\_burbano@berkeley.edu}
\affiliation{Perimeter Institute for Theoretical Physics, 
31 Caroline Street North, Waterloo, ON N2L 2Y5, Canada}
\affiliation{Department of Physics and Astronomy, University of Waterloo,
200 University Avenue West, Waterloo, ON N2L 3G1, Canada}

\author{T. Rick Perche}
\email{trickperche@perimeterinstitute.ca}
\affiliation{Perimeter Institute for Theoretical Physics,
31 Caroline Street North, Waterloo, ON N2L 2Y5, Canada}
\affiliation{Department of Applied Mathematics, University of Waterloo,
200 University Avenue West, Waterloo, ON N2L 3G1, Canada}

\author{Bruno de S. L. Torres}
\email{bdesouzaleaotorres@perimeterinstitute.ca}
\affiliation{Perimeter Institute for Theoretical Physics,
31 Caroline Street North, Waterloo, ON N2L 2Y5, Canada}
\affiliation{Department of Physics and Astronomy, University of Waterloo,
200 University Avenue West, Waterloo, ON N2L 3G1, Canada}
\affiliation{Institute for Quantum Computing, University of Waterloo,
200 University Avenue West, Waterloo, ON N2L 3G1, Canada}

\begin{abstract}
    Particle detectors are an ubiquitous tool for probing quantum fields in the context of relativistic quantum information (RQI). We formulate the Unruh-DeWitt (UDW) particle detector model in terms of the path integral formalism. The formulation is able to recover the results of the model in general globally hyperbolic spacetimes and for arbitrary detector trajectories. Integrating out the detector's degrees of freedom yields a line defect that allows one to express the transition probability in terms of Feynman diagrams. Inspired by the light-matter interaction, we propose a gauge invariant detector model whose associated line defect is related to the derivative of a Wilson line. This is another instance where nonlocal operators in gauge theories can be interpreted as physical probes for \textcolor{black}{quantum fields}. 

\end{abstract}

\maketitle

\tableofcontents

\section{Introduction}

One of the most basic questions that can be posed about the formulation of a physical theory is how it allows one to obtain information about a given physical system---in other words, roughly speaking, how one extracts \emph{measurable} quantities from the mathematical objects employed in the description of the theory. In classical mechanics, this discussion is often overlooked due to the natural association between two perspectives of position and momentum: as abstract coordinates on phase space or as physical quantities intrinsically associated with the mechanical system. While the former are mathematical tools used in the description of the system, the latter are in principle accessible through measurements with arbitrarily high precision. In quantum theory, however, the situation is much more subtle: states are no longer associated with well-defined values for all physical observables, and there is a fundamental limitation that forbids---even in principle---the extraction of arbitrarily precise values for all observables simultaneously. This places new challenges to the way measurements should be understood in a quantum theory at a very fundamental level. Elementary introductions to nonrelativistic quantum mechanics usually get away with this problem by postulating the idealized notion of projective measurements. However, this notion becomes untenable in quantum field theory (QFT) due to the extra requirements of relativistic locality and causality, which are in general violated by projective measurements acting on field states~\cite{sorkin_1956,fewster1,fewster2,fewster3}.

One efficient way to tackle the conceptualization of measurements in QFT is achieved with the aid of \emph{particle detector models}. In a few words, particle detectors are localized probes that couple to a quantum field in a local region of spacetime in a controllable fashion. They are usually thought of as effectively described by nonrelativistic quantum systems, so that one can avoid (at least on a practical level) the most fundamental obstructions to the implementation of measurements to the probe itself. Particle detectors are naturally appealing from a pragmatic point of view---after all, for all practical purposes, measurements of fields are always performed by coupling some probe to it. From a more theoretical perspective, they have also recently become the bread-and-butter of the field of RQI, where they are extensively used as the main tool to investigate a wide array of phenomena associated with the interplay between quantum information theory and relativity. This reinforces the importance of particle detectors as a useful theoretical tool to understand fundamental aspects in QFT.

In the context of RQI, the formulation of particle detector models in QFT is most commonly phrased in the framework of standard canonical quantization, with the interaction between detector and field usually prescribed at the level of a Hamiltonian. From such Hamiltonian, the time-evolution operator---and eventually all observables derived from the interaction---are then obtained. In QFT, however, path integral quantization provides a useful complementary point of view to the canonical methods. Being formulated directly in terms of the Lagrangian, the path integral approach is better adapted at keeping the symmetries of the theory manifest, while also providing a natural formulation in curved spacetimes. The path integral also yields a more direct route to \textcolor{black}{the construction of} correlation functions, \textcolor{black}{especially in perturbative treatments of interacting field theories}, and is extremely powerful for understanding the systematics of renormalization, as made explicit in the Wilsonian approach to the renormalization group. 

The ubiquity of the path integral as a fundamental tool in quantum field theory, together with the importance of particle detector models for the investigation of foundational aspects of quantum fields, motivates an approach to particle detectors via path integrals, which is what we address in this paper. Section~\ref{secUDW} reviews the most widely used particle detector model in RQI (the Unruh-DeWitt model), and describes how one can compute the most basic detector observable---namely, a vacuum excitation probability---via the usual canonical quantization framework. Section~\ref{secPI} then proceeds to introduce the description of the UDW model in terms of a path integral formulation; in particular, we show how the previous vacuum excitation probability is recovered from the path integral method. In Section~\ref{secLine}, we show how one can integrate out the detector degrees of freedom introduced in the previous section in order to generate an effective non-local field observable---a line defect---supported along the detector's worldline. With this, excitation amplitudes can be systematically computed in a perturbative expansion in terms of Feynman diagrams. 
In Section~\ref{secTopology}, we discuss how the case of a degenerate detector (when the energy gap between the two basis states vanishes) reduces the line defect to an expression similar to that of a Wilson line. This is in close analogy to the generally well-known fact that Wilson lines in non-Abelian gauge theories can be represented in the path integral as the effect of integrating out the spin degrees of freedom of external \textcolor{black}{quarks}~\cite{Tong2014}. \textcolor{black}{We further sharpen this connection by considering an effective model for the light-matter interaction where a nonrelativistic two-level atom interacts with the electromagnetic field via a dipole coupling. In that case, the line defect can be given a simple expression directly in terms of the \emph{derivative} of the Wilson line in the direction of the atom's dipole moment.} \textcolor{black}{In section \ref{sec:massive} we comment on the extension of this relation to detectors with an energy gap. We find that in perturbation theory the energy gap can be treated as a junction between the gapless defects previously found.} Overall, the connection between Wilson lines---and, more generally, line defects---with the effect of local probes coupled to quantum fields opens up the possibility of new venues for particle detector models. Indeed, while these models are most commonly restricted to real scalar fields, one can also explore them in the context of gauge theories, with potential applications to high energy and condensed matter physics.

\section{The UDW Model}\label{secUDW}
    
    Particle detector models are an elegant and simple way to probe quantum field theories. These can be used not only to probe fundamental aspects of QFT, such as the Unruh effect and Hawking radiation \cite{Takagi,Schlicht,ScullyPage,Unruh1976,Louko,Wald2,Unruh-Wald}, but also to probe the entanglement structure of the states associated with a quantum field \cite{Petar, Pozas-Kerstjens:2015, Pozas2016, Henderson2018, hottaHarvesting}. On top of that, numerous physical systems can be well modelled by particle detectors, such as atoms interacting with light \cite{Pozas2016,Nicho1,richard} or gravity \cite{remi}, and nucleons that decay via the weak force \cite{neutrinoOscillations}. 

    The simplest and most famous model of a particle detector is the UDW model \cite{Unruh1976,DeWitt}. It consists of a two-level quantum system that interacts linearly with a quantum scalar field, and can potentially be used in curved spacetimes \cite{us,Kukita2017,Henderson2019,Ng1,Ng2}. Although the smeared version of the UDW detector can be appealing for physical reasons \cite{us,us2,eduardo}, it has recently been shown that these are only covariant to lowest order in perturbation theory, and therefore cannot be used in fundamental descriptions \cite{us2}. On the other hand, the pointlike version of the UDW model has been shown to be fully covariant and consistent with causality. For this reason, we will work with the pointlike detector in this manuscript.
    
    \textcolor{black}{Let us begin by describing in detail the UDW model in its usual setting. Let us assume that the detector undergoes a timelike trajectory $z(\tau)$ parametrized by proper time $\tau$ in a given $D=n+1$ dimensional Lorentzian spacetime $\M$.} We \textcolor{black}{further} assume $\hat{\phi}(x)$ to be a free real scalar quantum field that can be written as a mode expansion
    \begin{equation}
        \hat{\phi}(x) = \int \dd[n]{\textcolor{black}{\bm{k}}} \left(u_{\bm k}(x) \hat{a}_{\bm k}+u_{\bm k}^*(x) \hat{a}^\dagger_{\bm k}\right),
    \end{equation}
    where $\{u_{\bm k}(x)\}$ is assumed to be a normalized basis of solutions to {the Klein-Gordon equation} and $(\hat{a}_{\bm k},\hat{a}_{\bm k}^\dagger)$ are the annihilation and creation operators {satisfying} the usual bosonic commutation relations. These operators allow one to build a Hilbert space representation for the quantum field, defined in terms of the vacuum state $\ket{0}$.
    
    The detector is prescribed as a two-level quantum system defined along $z(\tau)$. We assume its free Hamiltonian to be
    \begin{equation}\label{HD}
        \hat{H}_D = \frac{\Omega}{2}\left(\hat{\sigma}^+\hat{\sigma}^- -\hat{\sigma}^-\hat{\sigma}^+\right),
    \end{equation}
    where $\hat{\sigma}^+$ and $\hat{\sigma}^-$ are the $SU(2)$ ladder operators and $\Omega$ is the energy gap of the detector. From this formulation we define the ground and excited states of the detector, $\ket{g}$ and $\ket{e}$, by $\ket{e} = \hat{\sigma}^+\ket{g}$ and $\hat{\sigma}^-\ket{g} = 0$. It is important to remark that the free Hamiltonian $\hat{H}_D$ of the detector, given in the expression above, generates time translations with respect to the detector's proper time $\tau$. Indeed, if one were to compare the time evolution generated by $\hat{H}_D$ with the one generated by another notion of time translation $t$, one would obtain a redshift factor
    \begin{equation}
        \hat{H}_D^{(t)} = \ii \dv{}{t} = \ii \dv{\tau}{t}\dv{\tau} = \dv{\tau}{t}\hat{H}_D = \dv{\tau}{t}\frac{\Omega}{2}\left(\hat{\sigma}^+\hat{\sigma}^--\hat{\sigma}^-\hat{\sigma}^+\right).
    \end{equation}
    It is then arguable that the energy gap of the detector should transform as
    \begin{equation}
        \Omega \longmapsto \Omega \dv{\tau}{t}
    \end{equation}
    under time reparametrizations. Therefore, one could then see $\Omega$ as a 1-form defined along the detector's worldline $z(\tau)$. This viewpoint will be important once we discuss the geometrical perspective on particle detector models in Section \ref{secTopology}.
    
    The interaction of the detector with the quantum field is assumed to be linear on the field and can be prescribed by the following interaction Hamiltonian,
    \begin{equation}\label{eq:interation_UDW}
        \hat{H}_I(\tau) = \lambda \chi(\tau)(\hat{\sigma}^-(\tau) + \hat{\sigma}^+(\tau))\hat{\phi}(z(\tau)),
    \end{equation}
    where $\chi(\tau)$ is a switching function. We also assume the expression above to be written in the interaction picture, such that the operators involved evolve according to the free Hamiltonian.
    
    To obtain the transition probabilities associated with the detector, one usually approaches the model with a perturbative analysis, where the time evolution operator is given by the Dyson expansion
    \begin{align}
        \hat{U}_I &= \mathcal{T}\exp\left(-\ii \int \dd \tau \hat{H}_I(\tau)\right)
        = \sum_{n=0}^\infty \underbrace{\frac{(-\ii)^n}{n!}\int \dd\tau_1\cdots\dd\tau_n \mathcal{T}\Big(\hat{H}_I(\tau_1)\cdots\hat{H}_I(\tau_n)\Big)}_{\displaystyle{\hat{U}_I^{(n)}}}.
    \end{align}
    With the aid of the expansion above, it is possible to calculate the transition amplitudes associated with the excitation of the detector. There are multiple equivalent ways of performing this calculation. The simplest one is by considering the probability amplitude associated with the transition from the initial state $\ket{g,0}$ to a state \textcolor{black}{$\ket{e,\varphi}$}, where $\ket{\varphi}$ denotes an arbitrary final state of the field. We thus obtain
    \begin{align}\label{amplitudephi}
        &\mathcal{A}_{g\rightarrow e}(\varphi)=\bra{e,\varphi}\hat{U}_I\ket{g,0} = \sum_{n=0}^\infty \bra{e,\varphi}\hat{U}_I^{(n)}\ket{g,0}\\=& \sum_{n\textrm{ odd}}\lambda^{n}\frac{(-\ii)^{n}}{n!}\int \dd{\tau_1}\cdots\dd{\tau_{n}} \chi(\tau_1)\cdots\chi(\tau_{n}) \bra{\varphi}\mathcal{T}\Big(\hat{\phi}(\tau_1)\cdots\hat{\phi}(\tau_{n})\Big)\ket{0}e^{i\Omega(\tau_1-\tau_2+\dots+ \tau_{n})},\nonumber
    \end{align}
    where we have abbreviated $\hat{\phi}(z(\tau)) = \hat{\phi}(\tau)$. Notice that only the odd terms remain due to the fact that the transition probability is associated with the states $\ket{e}$ and $\ket{g}$ and the interaction is linear in $\hat{\sigma}^{+}$ and $\hat{\sigma}^{-}$. With this, it is possible to calculate the transition probability to arbitrary order by tracing over the field final states $\ket{\varphi}$, that is
    \begin{equation}\label{eq:excitation}
        \begin{aligned}
        p_{g\rightarrow e} &= \int \textrm{D}\varphi\, \abs{\mathcal{A}_{g\rightarrow e}(\varphi)}^2\\ &= \sum_{n,m\textrm{ odd}}\lambda^{n+m} \frac{ (-i)^{n-m}}{n!m!}\int \dd{\tau_1'}\dots\dd{\tau_{m}'} \dd{\tau_1}\cdots\dd{\tau_{n}} \chi(\tau_1')\cdots\chi(\tau_{m}') \chi(\tau_1)\cdots\chi(\tau_{n})\\&\hphantom{{}={}}\times \bra{0}\mathcal{T}\Big(\hat{\phi}(\tau_1')\cdots\hat{\phi}(\tau_{m}')\Big)^\dagger\mathcal{T}\Big(\hat{\phi}(\tau_1)\cdots\hat{\phi}(\tau_{n})\Big)\ket{0}e^{-i\Omega(\tau_1'-\dots+ \tau_{m}')}e^{i\Omega(\tau_1-\dots+ \tau_{n})},
        \end{aligned}
    \end{equation}
    where we have used the {resolution} of the identity in terms of field states $\ket{\varphi}$. In particular, the lowest order term is of second order in the coupling constant $\lambda$ and is given by
    \begin{equation}
        p_{g\rightarrow e}^{(2)} = \lambda^2 \int \dd\tau \dd\tau' \chi(\tau)\chi(\tau') \bra{0}\hat{\phi}(\tau)\hat{\phi}(\tau')\ket{0} e^{i\Omega(\tau-\tau')}.
    \end{equation}
    This is the most elementary detector observable usually considered in RQI that can be used to study phenomena such as the Unruh effect and Hawking radiation.
    
\section{Path Integral Formulation}\label{secPI}

    The goal of this section is to present a path integral formulation for the UDW model discussed in Section \ref{secUDW}. In order to do so we must associate classical actions to both the quantum field and the detector, so that the expectation values of observables can be computed using the standard techniques of the formalism. While the classical version of the quantum field is clear, when handling quantum systems with anti-commuting creation and annihilation operators through a path integral approach, one must introduce Grassmann variables and associate these to the states of the theory. In the UDW model presented in the previous section we had a two-level system. This can be modelled within the path integral formalism by introducing two Grassmann valued \textcolor{black}{fields} $\theta$ and $\bar{\theta}$ on the worldline of the detector \cite[see][Chapter 7]{Zinn-Justin2004}. However, in order to write the Hamiltonian \eqref{HD}, we must also ensure that our fields constitute a representation of $SU(2)$ after quantization. This is however impossible with only two Grassmann variables, which lead to the \textcolor{black}{Clifford} operator algebra $\text{Cl}_2$
    \begin{equation}\label{eq:car}
        \qty{\hat{\theta},\hat{\bar{\theta}}}=1\qand\qty{\hat{\theta},\hat{\theta}}=\qty{\hat{\bar{\theta}},\hat{\bar{\theta}}}=0.
    \end{equation}
    Instead, we must look for a set of Grassmann variables whose quantization yields an operator algebra $\text{Cl}_3$, which contains $\text{Spin}(3)\cong SU(2)$. Recalling that $\text{Cl}_3$ coincides with the even part of $\text{Cl}_4$ \cite[see][Theorem 3.7]{Lawson1989}, we can do this by considering the operators constructed with an even number of fields coming from two tuplets $(\theta,\bar{\theta})$ and $(\eta,\bar{\eta})$, whose quantizations form two anticommuting copies of \eqref{eq:car}. The model obtained has been used in \cite{Merad2001,Aouachria2002,Aouachria2009,Aouachria2013} to study the behaviour of spin 1/2 particles in a magnetic field.

    One way of proceeding is to write the ground and excited states of the UDW model as fermionic creation operators acting in a spurious vacuum state $\ket{0_D}$. That is, the goal is to write
    \begin{align}
        \ket{g}&=\hat{a}_g^\dagger \ket{0_D},\\
        \ket{e}&=\hat{a}_e^\dagger \ket{0_D},
    \end{align}
    where \textcolor{black}{the creation and annihilation operators $\hat{a}_i$ and $\hat{a}_i^\dagger$ satisfy the anti-commutation relations}
    \begin{align}
        \{\hat{a}_i,\hat{a}_j\} = 0, &&
        \{\hat{a}_i^\dagger,\hat{a}_j\} = \delta_{ij}, &&
        \{\hat{a}_i^\dagger,\hat{a}_j^\dagger\} = 0,
    \end{align}
    for $i,j=g,e$. This assigns the interpretation of the state $\ket{0_D}$ as a quantum state that represents the absence of the detector, while $\hat{a}_g^\dagger$ and $\hat{a}_e^\dagger$ can be seen as the operators that create the ground and excited states, respectively. One possible idea the reader may keep in mind is that the UDW detector is an atom, where the ground and excited states are associated with electrons in different shells. In this case, $\ket{0_D}$ would be associated with the vacuum of the electron field, while the $\hat{a}_i^\dagger$ and $\hat{a}_j$ operators would be the creation and annihilation operators in each of the shells. A similar reasoning has also been realized in~\cite{neutrinoOscillations}, where proton and neutron were interpreted as the ground and excited states of a detector that coupled to neutrinos via the weak force, and the ``true'' vacuum corresponded to the absence of nucleons.
    
    In terms of this description, the $\hat{\sigma}^+$ and $\hat{\sigma}^-$ operators that communicate between the ground and excited states can then be written as
    \begin{align}
        \hat{\sigma}^+ &= \hat{a}_e^\dagger\hat{a}_g,\\
        \hat{\sigma}^- &= \hat{a}_g^\dagger\hat{a}_e,
    \end{align}
    so that the free and interaction Hamiltonians of the system can be recast as
    \begin{equation}
        \begin{gathered}
        \hat{H}_D = \frac{\Omega}{2}\left( \hat{a}_e^\dagger \hat{a}_e- \hat{a}_g^\dagger \hat{a}_g\right),\\
        \hat{H}_I = \lambda \chi(\tau) \hat{\phi}(z(\tau))\left(\hat{a}_e^\dagger(\tau)\hat{a}_g(\tau)+\hat{a}_g^\dagger(\tau)\hat{a}_e(\tau)\right).
        \end{gathered}
    \end{equation}
    This description then allows one to associate the operators $\hat{a}_e^\dagger,\hat{a}_g^\dagger,\hat{a}_e,\hat{a}_g$ with the Grassmann variables $\bar{\theta},\bar{\eta},\theta,\eta$, respectively. Notice how the adjoint in the Hilbert space representation gets mapped into the Grassmann conjugation.
    
    With this rewriting of the UDW model in terms of the ground and excited states one can associate the following classical action for the full system consisting of the UDW detector, field and interaction
    \begin{equation}
        S = S_\phi + S_D + S_I,
    \end{equation}
    where $S_\phi$ is the action for the minimally coupled real scalar field $\phi$, $S_D$ is the action for the Grassmann variables  $\bar{\theta},\bar{\eta},\theta,\eta$ that describe the detector, and $S_I$ is the action associated with the interaction between them. These can be explicitly written as
    \begin{align}
        S_\phi &= \int \dd V\:\left(-\frac{1}{2}\nabla^\mu\phi \nabla_\mu\phi -\frac{m^2}{2} \phi^2\right),\\
        S_D &= \int \dd \tau \Bigg(\frac{\ii}{2}\left(\bar{\eta}\dot{\eta} - \dot{\bar{\eta}}\eta\right)+\frac{\ii}{2}(\bar{\theta}\dot{\theta} - \dot{\bar{\theta}}\theta)-\frac{\Omega}{2}\left( \bar{\theta}\theta- \bar{\eta}\eta\right)\Bigg),\label{actionDetector}\\
        S_I &= -\lambda\int \dd \tau \chi(\tau)\phi(z(\tau))(\theta\bar{\eta} + \eta\bar{\theta}).
    \end{align}   
    
    In summary, we assume the detector to be described on a classical level by Grassmann fields $\theta(\tau)$, $\eta(\tau)$, $\bar{\theta}(\tau)$ and $\bar{\eta}(\tau)$ defined on top of its predetermined trajectory $z(\tau)$. The equations of motion for these can be obtained by extremizing the action in Eq. \eqref{actionDetector}, and read
    \begin{align}
        i\dv{}{\tau}\theta &=\frac{\Omega}{2} \theta  ,\\
        i\dv{}{\tau}\eta &=-\frac{\Omega}{2} \eta .
    \end{align}
    Notice how the equations above are precisely the equations of motion satisfied by the quantum operators $\hat{a}_e$ and $\hat{a}_g$, respectively. This shows that the action in Eq. \eqref{actionDetector} is indeed appropriate for the description of the detector.

    Having this classical formulation for both the field and the detector, it is possible to apply the standard techniques of path integration to compute expectation values of observables. In order to calculate these and ensure that the integrals that arise are well defined, we perform a Wick rotation,
    \begin{equation}
        \tau \longmapsto -i \tau.
    \end{equation}
    We can then rewrite the action for the detector and the interaction with the quantum field as
\begin{align}
    S_D&=\int\dd{\tau}\qty(\bar{\theta}\dot{\theta}+\bar{\eta}\dot{\eta}+\frac{\Omega}{2}(\bar{\theta}\theta-\bar{\eta}\eta)),\label{eq:action_detector_real}\\
    S_I&=\lambda\int\dd{\tau}\chi(\tau)\phi(z(\tau))(\bar{\theta}\eta+\bar{\eta}\theta),\label{eq:action_interaction}
\end{align}
\textcolor{black}{where we have also integrated the first two terms in \eqref{actionDetector} by parts}. We will further assume that the support of the switching function $\chi(\tau)$ is compact and contained in an interval $[\tau_i,\tau_f]$ so that the boundary conditions in the integrals can be handled in a simple fashion. Let $\Sigma_i$ and $\Sigma_f$ be Cauchy hypersurfaces such that $z(\tau_i)\in\Sigma_i$ and $z(\tau_f)\in\Sigma_f$, and let $M$ be the region of spacetime bounded by these. In order to perform the path integration, let us choose classical field configurations $\varphi_i\in C^\infty(\Sigma_i)$ and $\varphi_f\in C^\infty(\Sigma_f)$. These can be associated with states $\ket{\varphi_i}$ and $\ket{\varphi_f}$ in the surfaces $\Sigma_i$ and $\Sigma_f$ so that $\hat{\phi}(x) \ket{\varphi_i} = \varphi_i(x) \ket{\varphi_i}$ for $x \in \Sigma_i$, and $\hat{\phi}(x) \ket{\varphi_f} = \varphi_f(x) \ket{\varphi_f}$ for $x \in \Sigma_f$. The path integral in this context will then be taken with respect to the boundary conditions $\theta(\tau_i)=\eta(\tau_i)=\bar{\theta}(\tau_f)=\bar{\eta}(\tau_f)=0$, $\phi|_{\Sigma_i}=\varphi_i$ and $\phi|_{\Sigma_f}=\varphi_f$. This will ensure that we will obtain the probability amplitudes of the form $\bra{0_D,\varphi_f}\hat{O}\ket{0_D,\varphi_i}$ whenever we integrate the classical observable associated with the operator $\hat{O}$. These boundary conditions also explain why we have omitted the boundary term obtained when going from \eqref{actionDetector} to \eqref{eq:action_detector_real}.

With this formalism at hand it is then possible to rewrite the transition probability amplitude for the field to start in the state $\ket{\varphi_i}$ and end in $\ket{\varphi_f}$ as
\begin{align}\label{amplitudephiif}
    \mathcal{A}_{g\rightarrow e}(\varphi_i,\varphi_f) = \ip{e,\varphi_f}{g,\varphi_i} = \mel{0_D,\varphi_f}{\hat{a}_e(\tau_f)\hat{a}_g^\dagger(\tau_i)}{0_D,\varphi_i}.
\end{align}
Recall that the classical insertions that play the role of the operators $\hat{a}_e(\tau_f)$ and $\hat{a}_g^\dagger(\tau_i)$ that show up in Eq. \eqref{amplitudephiif}  are $\theta(\tau_f)$ and $\bar{\eta}(\tau_i)$, respectively. With these we can then write the path integral that allows one, for instance, to calculate this transition amplitude
\begin{equation}\label{transitionamplitude}
    \mel{e,\varphi_f}{\hat{U}_I}{g,\varphi_i}=\int\mathcal{D}\phi\,\mathcal{D}\bar{\theta}\,\mathcal{D}\theta\,\mathcal{D}\bar{\eta}\,\mathcal{D}\eta\, e^{-S_\phi}e^{-S_D}e^{-S_I}\theta(\tau_f)\bar{\eta}(\tau_i).
\end{equation}
In order to simplify upcoming computations, we define 
\begin{equation}
    \xi = \begin{pmatrix}
        \eta\\
        \theta
    \end{pmatrix},\quad\quad\quad \bar{\xi} = \begin{pmatrix}
        \bar{\eta} &
        \bar{\theta}
    \end{pmatrix}, \quad\quad\quad \textrm{and}\quad\quad\quad \mathcal{D}\bar{\xi}\,\mathcal{D}\xi = \mathcal{D}\bar{\theta}\,\mathcal{D}\theta\,\mathcal{D}\bar{\eta}\,\mathcal{D}\eta.
\end{equation}

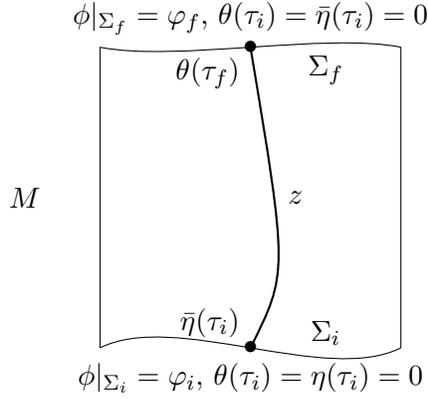
\begin{figure}
    \centering
    \begin{tikzpicture}
\draw (0,0) .. controls (1,0.5) and (3,-0.5) .. (4,0) ;
\draw (0,4) .. controls (1,3.8) and (3,4.2) .. (4,4) ;
\draw (0,0) -- (0,4);
\draw (4,0) -- (4,4);
\node at (-1,2) {$M$};
\draw[thick] (2,0) node {\textbullet} node[anchor = south east] {$\bar{\eta}(\tau_i)$} .. controls (2.5,1) .. (2,4) node {\textbullet} node[anchor = north east] {$\theta(\tau_f)$};
\node at (2.6,2) {$z$};
\node at (3,0.2) {$\Sigma_i$};
\node at (3,3.7) {$\Sigma_f$};
\node at (2,4.4) {$\phi|_{\Sigma_f}=\varphi_f$, $\bar{\theta}(\tau_i)=\bar{\eta}(\tau_i)=0$};
\node at (2,-0.4) {{$\phi|_{\Sigma_i}=\varphi_i$, $\theta(\tau_i)=\eta(\tau_i)=0$}}; 
\end{tikzpicture}
\caption{Path integral for the excitation amplitude of the detector between two arbitrary field configurations at different Cauchy hypersurfaces. The region enclosed by these is denoted by $M$. Such a region defines a time evolution operator $U(M)$ between the Cauchy hypersurfaces.\label{fig:path_integral}}
\end{figure}

Let us show how to use this formalism to recover the results from Section \ref{secUDW}. In a perturbative analysis, it is possible to expand the interaction action $S_I$ in powers of $\lambda$ so that one obtains
\begin{align}
    &\ip{e,\varphi_f}{g,\varphi_i}=\sum_{n=0}^\infty \frac{(-1)^n}{n!} \int\mathcal{D}\phi\,\mathcal{D}\bar{\xi}\,\mathcal{D}\xi\, e^{-S_\phi}e^{-S_D}(S_I)^n\theta(\tau_f)\bar{\eta}(\tau_i)\\
    &=\sum_{n=0}^\infty\frac{(-1)^n}{n!}\int\mathcal{D}\phi\,\mathcal{D}\bar{\xi}\,\mathcal{D}\xi\, e^{-S_\phi}e^{-S_D} \theta(\tau_f)\left(\lambda\int \dd \tau \chi(\tau)\phi(\tau)(\bar{\theta}(\tau)\eta(\tau)+\bar{\eta}(\tau)\theta(\tau))\right)^n\bar{\eta}(\tau_i)\nonumber.
\end{align}
Before analyzing the general expansion, notice that the zeroth order term yields zero (as expected) because $S_D$ does not couple $\theta$ and $\bar{\eta}$. The first order transition amplitude can then be written as
\begin{equation}
    \begin{aligned}
    \mathcal{A}&_{g\rightarrow e}^{(1)}(\varphi_i,\varphi_f) \\&=-\lambda \int\mathcal{D}\phi\,\mathcal{D}\bar{\xi}\,\mathcal{D}\xi\, e^{-S_\phi}e^{-S_D} \theta(\tau_f)\int \dd \tau \chi(\tau)\phi(\tau)(\bar{\theta}(\tau)\eta(\tau)+\bar{\eta}(\tau)\theta(\tau))\bar{\eta}(\tau_i)\\
    &=-\lambda\int \dd \tau\chi(\tau)\int\mathcal{D}\phi e^{-S_\phi}\phi(\tau)\,\mathcal{D}\bar{\xi}\,\mathcal{D}\xi\, e^{-S_D} \theta(\tau_f) (\bar{\theta}(\tau)\eta(\tau)+\bar{\eta}(\tau)\theta(\tau))\bar{\eta}(\tau_i)\\
    &=-\lambda\int \dd \tau\chi(\tau)\bra{\varphi_f}\hat{\phi}(\tau)\ket{\varphi_i}\bra{0_D} \hat{a}_e(\tau_f) \hat{a}_e^\dagger(\tau)\hat{a}_g(\tau)\hat{a}_g^\dagger(\tau_i)\ket{0_D}\\
    &=-\lambda e^{-\frac{\Omega}{2}(\tau_i+\tau_f)}\int \dd \tau\chi(\tau)\bra{\varphi_f}\hat{\phi}(\tau)\ket{\varphi_i}e^{\Omega\tau},
    \end{aligned}
\end{equation}
where we have used the fact that the boundary conditions on the $\phi$ integral imply that the integral above yields the matrix element $\bra{\varphi_f}\hat{\phi}(x)\ket{\varphi_i}$. This means that we obtain the exact analogue of the first order term of Eq. \eqref{amplitudephi}, except for a Wick rotated global phase $ e^{-\frac{\Omega}{2}(\tau_i+\tau_f)}$. Of course this preliminary result can then be used to recover the vacuum transition probability by integrating over the field final states and decomposing the vacuum in terms of the $\ket{\varphi}$ states. But instead of performing the calculations using this line of thought, it is easier to deduce the Feynman rules associated with this interacting theory. We will do so in the next section, where we will interpret the detector as \textcolor{black}{a line defect}.
    
\section{The Detector as a Line Defect}\label{secLine}

The path integral formulation presented in the previous section yields a convenient formalism which allows us to integrate out the degrees of freedom of the detector. Indeed, in Eq. \eqref{transitionamplitude}, we can extract the detector's dependence as an insertion $D_z(\phi)$ in the field path integral. It is given by
\begin{equation}\label{eq:line_defect}
    D_z(\phi)=\int\mathcal{D}\bar{\theta}\,\mathcal{D}\theta\,\mathcal{D}\bar{\eta}\,\mathcal{D}\eta\, e^{-S_D-S_I}\theta(\tau_f)\bar{\eta}(\tau_i).
\end{equation}
This leaves us with an effective operator supported along the worldline $z$ such that the excitation amplitudes will be given completely in terms of our quantum field
\begin{equation}
    \ip{e,\varphi_f}{g,\varphi_i}=\int\mathcal{D}\phi\, e^{-S_\phi}D_z(\phi)=\mel{\varphi_f}{\mathcal{T}D_z(\hat{\phi})}{\varphi_i}.
\end{equation}
Operators of this form---non-local operators supported along a line on spacetime---are generically known as line defects \cite{queTrampoSpringer}.

At this point it is interesting to remark that this is an implementation of the ideas featured in \cite{fewster1,fewster2,fewster3}. Indeed, with \eqref{eq:line_defect} we have found the field observable corresponding to the presence of the detector.

The line defect \eqref{eq:line_defect} can be computed in terms of Feynman diagrams. The free action $S_D$ yields the propagators
\begin{equation}\label{eq:propagator}
    \begin{aligned}
\begin{tikzpicture}[baseline]
\begin{feynman}
\vertex (in) at (-2,0) {$\tau_1$};
\vertex (out) at (2,0) {$\tau_2$};
\diagram*{
(in) -- [fermion] (out)
};
\end{feynman}
\end{tikzpicture}&= \ev{\theta(\tau_2)\bar{\theta}(\tau_1)}=e^{-\frac{\Omega}{2}(\tau_2-\tau_1)}\Theta(\tau_2-\tau_1),\\
\begin{tikzpicture}[baseline]
\begin{feynman}
\vertex (in) at (-2,0) {$\tau_1$};
\vertex (out) at (2,0) {$\tau_2$};
\diagram*{
(in) -- [charged scalar] (out)
};
\end{feynman}
\end{tikzpicture}&= \ev{\eta(\tau_2)\bar{\eta}(\tau_1)}=e^{\frac{\Omega}{2}(\tau_2-\tau_1)}\Theta(\tau_2-\tau_1),
\end{aligned}
\end{equation}
where $\ev{\cdot}$ denotes the path integral over the fermionic variables in $[\tau_i,\tau_f]$ with the boundary conditions prescribed above. Indeed, one can easily check that the boundary conditions $\ev{\theta(\tau_i)\bar{\theta}(\tau_1)}=\ev{\theta(\tau_2)\bar{\theta}(\tau_f)}=0$ are satisfied \textcolor{black}{so that} we have a Green's function
\begin{equation}
    \qty(\dv{\tau_2}+\frac{\Omega}{2})\ev{\theta(\tau_2)\bar{\theta}(\tau_1)}=\delta(\tau_2-\tau_1),
\end{equation}
and the analogous expression holds for $\eta(\tau)$ and $\bar{\eta}(\tau)$. The cubic term $S_I$ in the action yields the interaction vertices
\begin{equation}
\begin{tikzpicture}[baseline]
\begin{feynman}
\vertex (in) at (-2,0);
\vertex (phi) at (0,1);
\vertex[label = -90:\(\tau\)] (int) at (0,0);
\vertex (out) at (2,0);
\diagram*{
(in) -- [fermion] (int) -- [charged scalar] (out),
(phi) -- [gluon] (int)
};
\end{feynman}
\end{tikzpicture}=-\lambda\chi(\tau)\phi(\tau)=\begin{tikzpicture}[baseline]
\begin{feynman}
\vertex (in) at (-2,0);
\vertex (phi) at (0,1);
\vertex[label = -90:\(\tau\)] (int) at (0,0);
\vertex (out) at (2,0);
\diagram*{
(in) -- [charged scalar] (int) -- [fermion] (out),
(phi) -- [gluon] (int)
};
\end{feynman}
\end{tikzpicture}.
\end{equation}
With these we now have that $D_z(\phi)$ corresponds to diagrams of the form
\begin{equation}\label{eq:building_block}
    \begin{tikzpicture}[baseline]
    \begin{feynman}
    \vertex (in) at (-2,0) {\(\tau_i\)};
    \vertex[blob] (int) at (0,0) {};
    \vertex (out) at (2,0) {\(\tau_f\)};
    \diagram*{
    (in) -- [charged scalar] (int) -- [fermion] (out)
    };
    \end{feynman}
    \end{tikzpicture},
\end{equation}
where the external legs come from the operator insertions at the boundaries in \eqref{eq:line_defect}. Notice that time ordering prevents the appearance of vacuum diagrams. This corresponds to the fact that the path integral without insertions is 1, or equivalently the vacuum \textcolor{black}{state} $\ket{0_D}$ of our fermionic system is normalized. One might be worried that this prevents us from having loops but it is important to remember that these Feynman diagrams are in the worldline, not spacetime. In particular, nothing prevents $z(\tau_f)=z(\tau_i)$. Then, the order $\mathcal{O}(\lambda^n)$ term of $D_z(\phi)$ is
\begin{equation}\label{eq:UDW_defect}
\begin{aligned}
    &\hphantom{{}={}}\begin{tikzpicture}[baseline]
    \begin{feynman}
    \vertex (in) at (-6,0) {\(\tau_i\)};
    \vertex[label = -90:\(\tau_1\)] (int1) at (-4,0);
    \vertex (phi1) at (-4,1);
    \vertex[label = -90:\(\tau_2\)] (int2) at (-2,0);
    \vertex (phi2) at (-2,1);
    \vertex[label = -90:\(\tau_3\)] (int3) at (0,0);
    \vertex (phi3) at (0,1);
    \vertex[circle] (dots) at (2,0) {\(\cdots\)};
    \vertex[label = -90:\(\tau_n\)] (intn) at (4,0);
    \vertex (phin) at (4,1);
    \vertex (out) at (6,0) {\(\tau_f\)};
    \diagram*{
    (in) -- [charged scalar] (int1) -- [fermion] (int2) -- [charged scalar] (int3) -- [fermion] (dots) -- [charged scalar] (intn) -- [fermion] (out),
    (phi1) -- [gluon] (int1),
    (phi2) -- [gluon] (int2),
    (phi3) -- [gluon] (int3),
    (phin) -- [gluon] (intn)
    };
    \end{feynman}
    \end{tikzpicture}\\
    &=(-\lambda)^n\int_{\tau_i}^{\tau_f}\dd{\tau_1}{\int_{\tau_1}^{\tau_f}}\dd{\tau_2}\cdots\int_{\tau_{n-1}}^{\tau_f}\dd{\tau_n}\chi(\tau_n)\phi(\tau_n)\cdots\chi(\tau_1)\phi(\tau_1)\\
    &\hphantom{{}={}PHAAAAAAAAAANTOOOOM}\times e^{\frac{\Omega}{2}(\tau_1-\tau_i)}e^{-\frac{\Omega}{2}(\tau_2-\tau_1)}e^{\frac{\Omega}{2}(\tau_3-\tau_2)}\cdots e^{-\frac{\Omega}{2}(\tau_f-\tau_n)}\\
     &=(-\lambda)^ne^{-\frac{\Omega}{2}(\tau_i+\tau_f)}\int_{\tau_i}^{\tau_f}\dd{\tau_1}\cdots\int_{\tau_{n-1}}^{\tau_f}\dd{\tau_n}\chi(\tau_n)\phi(\tau_n)\cdots\chi(\tau_1)\phi(\tau_1)e^{\Omega(\tau_1-\tau_2+\tau_3-\cdots+\tau_n)}.
\end{aligned}
\end{equation}
The contributions for even $n$ vanish since these diagrams are only possible for $n$ odd, agreeing with the results from Section \ref{secUDW}.

As a first application of this line defect, let us recover the excitation probability \eqref{eq:excitation}. The excitation amplitude from the uncoupled field-detector vacuum state into an arbitrary state $\ket{\varphi_f}$ of the field can be written as
\begin{equation}
\begin{aligned}
    \ip{e,\varphi_f}{g,0}&=\int\textrm{D}\varphi\,\ip{e,\varphi_f}{g,\varphi}\ip{\varphi}{0}=\int\textrm{D}\varphi\,\mel{\varphi_f}{\mathcal{T}D_z(\hat{\phi})}{\varphi}\ip{\varphi}{0}\\
    &=\mel{\varphi_f}{\mathcal{T}D_z(\hat{\phi})}{0}
\end{aligned}
\end{equation}
by introducing the identity $\openone=\int\textrm{D}\varphi\,\dyad{\varphi}$ on the Hilbert space of the field at $\Sigma_i$. On the other hand, its conjugate is obtained within the path integral formalism by considering the manifold $M^-$, which coincides with $M$ but has incoming boundary $\Sigma_f$ and outgoing boundary $\Sigma_i$. We then extend the detector's trajectory and the switching function to $M^-$ in a natural way. Repeating the argument above one can see it is explicitly given by
\begin{equation}
    \ip{g,0}{e,\varphi_f}=\mel{0}{\mathcal{T}D_{z^-}(\hat{\phi})}{\varphi_f},
\end{equation}
where $z^-$ is the mirror trajectory in $M^-$ of the detector and $D_{z^-}(\hat{\phi})$ uses the mirrored switching function $\chi^-(\tau)$. For gluing purposes, we will assume this is parametrized on $[\tau_f,2\tau_f-\tau_i]$ so that $z^-(\tau) = z(2\tau_f-\tau)$ and $\chi^-(\tau) = \chi(2\tau_f-\tau)$. The full transition probability is then given by multiplying these two amplitudes together. The excitation probability of the detector, regardless of the final state of the field, is obtained by integrating over $\varphi_f$
\begin{equation}\label{eq:probability_gluing}
    p_{g\rightarrow e}=\int\textrm{D}\varphi_f\,\mel{0}{\mathcal{T}D_{z^-}(\hat{\phi})}{\varphi_f}\mel{\varphi_f}{\mathcal{T}D_z(\hat{\phi})}{0}=
    \mel{0}{\mathcal{T}D_{z^-}(\hat{\phi})\mathcal{T}D_z(\hat{\phi})}{0}.
\end{equation}
This corresponds to gluing $M$ and $M^-$ as shown in Figure \ref{fig:paste}. Our probability can \textcolor{black}{then} be written in terms of the composite trajectory $z^- z$ as
\begin{equation}\label{eq:probability_line_integral}
    p_{g\rightarrow e}=
    \ev{\mathcal{T}D_{z^ -z}(\hat{\phi})}{0}.
\end{equation}%
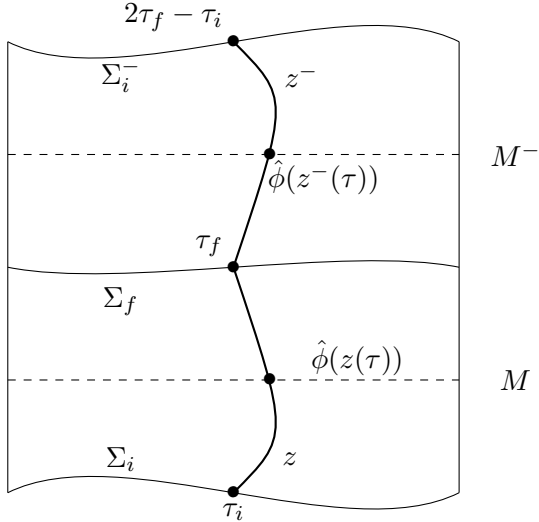
\begin{figure}
    \centering
\begin{tikzpicture}[scale = 1.5]
\draw (0,0) .. controls (1,0.5) and (3,-0.5) .. (4,0) ;
\draw (0,2) .. controls (1,1.8) and (3,2.2) .. (4,2) ;
\draw (0,4) .. controls (1,3.5) and (3,4.5) .. (4,4) ;
\draw (0,0) -- (0,4);
\draw (4,0) -- (4,4);
\node at (4.5,1) {$M$};
\node at (4.5,3) {$M^-$};
\draw[thick] (2,0) node {\textbullet} node[anchor = north] {$\tau_i$} .. controls (2.5,0.5) .. (2,2) node {\textbullet} node[anchor = south east] {$\tau_f$};
\draw[thick] (2,2) .. controls (2.5,3.5) .. (2,4) node {\textbullet} node[anchor = south east] {$2\tau_f-\tau_i$};
\node at (2.5,0.3) {$z$};
\node at (2.6,3.7) {$z^-$};
\node at (1,0.3) {$\Sigma_i$};
\node at (1,1.7) {$\Sigma_f$};
\node at (1,3.7) {$\Sigma_i^-$};
\node at (2.32,1) {\textbullet};
\node at (3.1,1.2) {$\hat{\phi}(z(\tau))$};
\draw[dashed] (0,1) -- (4,1);
\node at (2.32,3) {\textbullet};
\node at (2.8,2.8) {$\hat{\phi}(z^ -(\tau))$};
\draw[dashed] (0,3) -- (4,3);
\end{tikzpicture}
    \caption{Pasting of the integral in figure \ref{fig:path_integral} (after removing the fermionic degrees of freedom) along $\Sigma_f$
    \label{fig:paste}} 
\end{figure}%
From the expression of the line defect in terms of the path integral given in~\eqref{eq:line_defect}, we see that this constitutes a nonperturbative expression for the excitation probability, suitable for computations within the path integral.

Finally, let us comment on how one can see the connection between \eqref{eq:excitation} and \eqref{eq:probability_line_integral} directly. We defined the extension $z^-$ of our trajectory so that $\hat{\phi}(z^-(\tau))=\hat{\phi}(z(2\tau_f-\tau))$. Therefore, within the operator formalism we can reduce this probability by expressing the line defect on $z^-$ in terms of a new line defect on $z$. In doing this however, we must be careful with the interpretation of the time ordered product. For example, from figure \ref{fig:paste}, we can see that for $\tau_2>\tau_1$ the time ordered product is $\hat{\phi}(z^-(\tau_2))\hat{\phi}(z^-(\tau_1))=\hat{\phi}(z(2\tau_f-\tau_2))\hat{\phi}(z(2\tau_f-\tau_1))$. From the point of view of $z$ this would have the opposite ordering. With this in mind, after a change in the integration variables and taking into account the fact that the orientation in $M^-$ is reversed, the line defect $\mathcal{T}D_{z^-}(\hat{\phi})$ at $\mathcal{O}(\lambda^n)$ can be written as (suppressing the switching function)
\begin{align}
    &\hphantom{{}={}}(-\lambda)^n\int_{\tau_f}^{2\tau_f-\tau_i}(-\dd{\tau_1})\cdots\int_{\tau_{n-1}}^{2\tau_f-\tau_i}(-\dd{\tau_n})\hat{\phi}(z(2\tau_f-\tau_n))\cdots\hat{\phi}(z(2\tau_f-\tau_1))\nonumber\\
    &\hphantom{{}={}\lambda^m\int_{\tau_f}^{2\tau_f-\tau_i}\dd{\tau_1}\cdots\int_{\tau_{m-1}}^{2\tau_f-\tau_i}\dd{\tau_m}}e^{\frac{\Omega}{2}(\tau_1-\tau_f)}e^{-\frac{\Omega}{2}(\tau_2-\tau_1)}e^{\frac{\Omega}{2}(\tau_3-\tau_2)}\cdots e^{-\frac{\Omega}{2}(2\tau_f-\tau_i-\tau_m)}\nonumber\\
    &=\lambda^n\int_{\tau_i}^{\tau_f}\dd{\tau_1}\cdots\int_{\tau_i}^{\tau_{n-1}}\dd{\tau_n}\hat{\phi}(z(\tau_n))\cdots\hat{\phi}(z(\tau_1))\\
    &\hphantom{{}={}\lambda^m\int_{\tau_f}^{2\tau_f-\tau_i}\dd{\tau_1}\cdots\int_{\tau_{m-1}}^{2\tau_f-\tau_i}\dd{\tau_m}}e^{\frac{\Omega}{2}(\tau_f-\tau_1)}e^{-\frac{\Omega}{2}(\tau_1-\tau_2)}e^{\frac{\Omega}{2}(\tau_2-\tau_3)}\cdots e^{-\frac{\Omega}{2}(\tau_n-\tau_i)},\nonumber
\end{align}
which coincides with $(\mathcal{T} D_z(\hat{\phi}))^\dagger$ as can be seen by comparison with \eqref{eq:UDW_defect} after relabelling $\tau_j\rightarrow \tau_{n+1-j}$ and reorganizing the integration limits. We conclude that $p_{g\rightarrow e}=\ev{(\mathcal{T} D_z(\hat{\phi}))^\dagger\mathcal{T} D_z(\hat{\phi})}{0}$, which corresponds to \eqref{eq:excitation}.
\color{black}

\section{Gauge Particle Detector Models}\label{secTopology}

    Line defects play a major role in the study of gauge theories, where they are commonly known as Wilson lines. These defects have been shown to completely determine the gauge connection~\cite{ivaan} and thus are commonly considered to be the fundamental observables of these theories.
    In this section we 
    establish a relationship between particle detectors and Wilson lines. As expected \cite[see][Section 2.1.3]{Tong2018}, in the case $\Omega=0$, where the Hamiltonian of the detector vanishes and the detector itself is an instance of topological quantum mechanics, the line defect takes the form
    \begin{equation}\label{eq:sinh_defect}
        D_z(\phi)=\sinh(-\lambda\int\dd{\tau} \chi(\tau)\phi(z(\tau))),
    \end{equation}
    which is analogous to the Wilson lines one might expect from a gauge theory. The appearance of $\sinh$ instead of $\exp$ is due to the vanishing of the even orders in perturbation theory. These can be recovered, for example, by adding another line defect obtained by replacing the insertion $\theta(\tau_f)\bar{\eta}(\tau_i)$ in \eqref{eq:line_defect} by $\theta(\tau_f)\bar{\theta}(\tau_i)$. In particular, the path ordered exponential could be obtained as the line defect corresponding to the sum of the amplitudes associated with remaining in the ground state and transitioning to the excited state. Its square then measures the interference between these two possibilities.
    
    Usually, when one {refers to} the UDW model, the model presented in Section \ref{secUDW} is implied, where the detector interacts with a scalar field. However, the model can be motivated by the interaction of nonrelativistic systems with the electromagnetic field \cite{MariaGoeppert-Meyer,richard}. The form of \eqref{eq:sinh_defect} then suggests a connection between the UDW defect and the Wilson lines of the underlying electromagnetic theory. To see this, let us begin by revisiting the origin of the UDW detector from the interaction of a hydrogen-like atom with light. The interaction Hamiltonian between an electron in an atom with the electromagnetic field due to its electric dipole is given by
    \begin{equation}\label{eq:light-matter}
        \hat{H}_I= -e\int\dd[3]{\bm{x}}\hat{\bm{d}}(x)\cdot\hat{\bm{E}}(x),
    \end{equation}
    where $e$ stands for the electron's charge, which plays the role of the coupling constant $\lambda$ in this case, and $\hat{\bm{d}}(x)=\chi(\tau)\left(\bm{V}({x})\hat{\sigma}^+(\tau)+\bm{V}^*({x})
    \hat{\sigma}^-(\tau)\right)$ is the dipole operator of the atom restricted to two energy levels. Notice that we have also incorporated the switching function $\chi(\tau)$ in the definition of $\hat{\bm{d}}(x)$. Roughly speaking, the UDW model \eqref{eq:interation_UDW} can be motivated by replacing the dipole coupling $\hat{\bm{d}}(x)\cdot \hat{\bm{E}}(x)$ by a smeared scalar field $\Lambda(x)\hat{\phi}(x)$, where $\Lambda(x)$ is the spacetime smearing function \cite{us,us2}.
    
    However, it is possible to recast \eqref{eq:light-matter} in a covariant way that preserves the coupling with a vector field in the pointlike limit. Assume that the dipole can be written as \mbox{$\hat{\bm{d}}({x}) = f(\bm{x}) (\hat{\sigma}^+(\tau)+\hat{\sigma}^-(\tau))\bm{X}(x)$} for a real smearing function $f(\bm{x})$ and a vector field $\bm{X}$ that contains the time dependence of the interaction. We take a pointlike limit by assuming $f(\bm{x})$ to be a Dirac delta function and extend all the vectors to spacetime by considering an orthonormal frame such that the timelike vector is given by $\dot{z}^\mu(\tau)$. In this case, we can write $\bm{E}(\tau,\bm{x}) \to \bm{E}(z(\tau))$ as $\dot{z}^\mu F_{\mu\nu}(z)$ so that
    \begin{equation}
        \int\dd[3]{\bm{x}}\hat{\bm{d}}(\tau,\bm{x})\cdot\hat{\bm{E}}(\tau,\bm{x}) \longrightarrow  (\sigma^+(\tau) + \sigma^-(\tau))\dot{z}^\mu \hat{F}_{\mu\nu}(z) X^\nu.
    \end{equation}
    This, together with the resemblance between the UDW line defect and Wilson loops suggests an alternative to the interaction action \eqref{eq:action_interaction},
    \begin{equation}\label{thisActionHere}
        S_I= \lambda \int \dd{\tau}  \dot{z}^\mu F_{\mu\nu}(z)X^\nu(\bar{\theta}\eta+\bar{\eta}\theta),
    \end{equation}
    where the pullback of the scalar field has been replaced by the pullback of the 1-form $F_{\mu\nu} X^\nu$. This is a very natural choice, because the resulting action is gauge invariant, provided that the detector degrees of freedom do not transform under the action of the electromagnetic potential $A_\mu$. Moreover, the resulting action is now reparametrization invariant and covariant in general spacetimes. This detector model is then a geometrical model, for it can be expressed in terms of the geometry of the trajectory in spacetime without reference to a particular parametrization. In fact, the dependence of the model in the spacetime metric is now completely contained within the dynamics of the gauge field.
    
    There is an important remark regarding the gauge invariance of the model. Notice that in an atom one would have that the electron wavefunctions transforms under gauge transformations. However, to reach the dipole coupling \eqref{eq:light-matter}, one must gauge fix and perform the dipole approximation. These end up resulting in an emergent gauge symmetry for the model, where the internal degrees of freedom of the detector do not transform under gauge transformations.    
    
    The associated line defect of the model---which coincides with the excitation amplitude---can be obtained immediately via the replacement $\phi(z)\mapsto \dot{z}^\mu F_{\mu\nu}(z)X^\nu$ in \eqref{eq:UDW_defect}. In particular, in the limit $\Omega=0$, it can be written as
    \begin{equation}
        D_z(A) = \sinh\left(-\lambda\int\dd{\tau}\:\dot{z}^\mu F_{\mu\nu}(z)X^\nu\right).
    \end{equation}
    In this case, the line defect bears a very interesting relationship with the Wilson lines $W_z(A)$, defined by
    \begin{equation}
        W_z(A) = \exp\left(\:\:\int_z\:\: A\:\:\:\: \right),
    \end{equation}
    where $z$ represents a trajectory in spacetime. Indeed, it is possible to express the action \eqref{thisActionHere} in terms of the derivative of the Wilson line $W_z(A)$ when one shifts the trajectory by the flow of the vector field $X(\tau)$ defined along the worldline. More precisely, let $\Phi_\varepsilon$ denote \textcolor{black}{a flow by a parameter $\varepsilon$ agreeing with $X$ along the worldline}.  Then we have
    \begin{align}\label{eq:linedefectderivative}
        X\log W_z(A) = \lim_{\varepsilon \to 0} \frac{1}{\varepsilon}\left(\int_{\Phi_\varepsilon(z)}\!\!\! A-\int_z A \right) = \lim_{\varepsilon \to 0}\frac{1}{\varepsilon} \oint_{\partial\Sigma_{\varepsilon}}\!\!  A= \lim_{\varepsilon \to 0}\frac{1}{\varepsilon} \int_{\Sigma_\varepsilon} F  = \int \dd\tau  X^\mu \dot{z}^\nu F_{\mu\nu},
    \end{align}
    where $\Sigma_\varepsilon$ is the ribbon defined by $\Phi_s(z)$ for $s \in [0,\varepsilon]$ (Fig \eqref{fig:derivative_line}). \textcolor{black}{The details of this computation can be found in Appendix \ref{Bruno}}. In particular, in this case, the line defect $D_z(A)$ can be written in terms of the Wilson line according to
    \begin{equation}\label{eq:derivative_defect}
        D_z(A) = \sinh\left(\lambda X \log W_z(A)\right),
    \end{equation}
    making explicit its gauge independence and the relationship of the response of a gauge UDW detector with Wilson lines.
    
    There are natural generalizations of this model to the non-Abelian case whose corresponding defects exhibit an analogue of the relation \eqref{eq:derivative_defect} in the $\Omega=0$ limit. They are described by an action $S=S_{\text{gauge}}+S_D+S_I$, where $S_{\text{gauge}}$ is a gauge-invariant action for a gauge field $A$, $S_D$ is the detector action \eqref{eq:action_detector_real}, and
    \begin{equation}
        S_I=\lambda\int\dd{\tau}\dot{z}^\mu\Tr_R(F_{\mu\nu})X^\nu(\bar{\theta}\eta+\bar{\eta}\theta),
    \end{equation}
    for some representation $R$ of the gauge group and a vector field $X$ along $z$ vanishing at the endpoints. In particular, the gauge invariance of $S_{\text{gauge}}$ extends to the full model by declaring the transformation behaviour of the fermions to be trivial. 
    
    Much like in the Abelian case, the calculations of section \ref{secLine} remain valid under the identification $\phi\mapsto \dot{z}^\mu\Tr_R(F_{\mu\nu})X^\nu$. In particular, at $\Omega=0$ the defect corresponding to excitation amplitudes is of the form
    \begin{equation}
        D_z(A)=\sinh(-\lambda\int\dd{\tau}\dot{z}^\mu \Tr_R(F_{\mu\nu})X^\nu).
    \end{equation}
    The extension of equation \eqref{eq:derivative_defect} can be obtained by computing the derivative of the corresponding Wilson line
    \begin{equation}
        W_z(A)=\mathcal{P}\exp(\int_{z}A),
    \end{equation}
    which in the non-Abelian case requires the path ordering denoted by $\mathcal{P}$. In the case that $z$ is a loop, this derivative has been computed to be \cite[see][equation (4.78)]{Cherednikov2014}
    \begin{equation}\label{eq:non_abelian_derivative}
        XW_z(A)=W_z(A)\int\dd{\tau} W_{z^\tau}(A) X^\mu F_{\mu\nu}(z)\dot{z}^\nu W_{z^\tau}(A)^{-1},
    \end{equation}
    where $z^\tau$ is the restriction to $[\tau_i,\tau]$ of $z$. This result remains valid in our case since $X$ vanishes at the endpoints. Indeed, we can complete the trajectory $z$ to a loop by adding to it another trajectory $\tilde{z}$. Simultaneously, we will extend $X$ to the loop by setting it to $0$ along $\tilde{z}$, so that $X W_{\tilde{z}z}(A)=X (W_{\tilde{z}}(A)W_z(A))=W_{\tilde{z}}(A)XW_z(A)$. On the other hand, we have
    \begin{equation}
    \begin{aligned}
        X W_{\tilde{z}z}(A)&=W_{\tilde{z}z}(A)\int\dd{\tau} W_{({\tilde{z}z})^\tau}(A) X^\mu F_{\mu\nu}(z)\dot{z}^\nu W_{({\tilde{z}z})^\tau}(A)^{-1}\\
        &=W_{\tilde{z}}(A)W_z(A)\int\dd{\tau} W_{z^\tau}(A) X^\mu F_{\mu\nu}(z)\dot{z}^\nu W_{z^\tau}(A)^{-1},
    \end{aligned}
    \end{equation}
    where, given that $X$ vanishes on $\tilde{z}$, we have restricted the integral to $z$ and used the fact that on this restriction $(\tilde{z}z)^\tau=z^\tau$. Thus, cancelling the factors of $W_{\tilde{z}}(A)$ we recover \eqref{eq:non_abelian_derivative} for our possibly open path $z$. Taking traces of the logarithmic derivative then cancels the Wilson factors inside the integral so that we obtain the analogue of \eqref{eq:derivative_defect}  
    \begin{equation}\label{eq:derivative_defect_non_abelian}
        D_z(A)=\sinh(\lambda \Tr_R(X\log W_z(A))).
    \end{equation}
    
    \begin{figure}
        \centering
       \begin{tikzpicture}[scale = 1.5]
\draw (0,0) .. controls (1,0.5) and (3,-0.5) .. (4,0) ;
\draw (0,4) .. controls (1,3.5) and (3,4.5) .. (4,4) ;
\draw (0,0) -- (0,4);
\draw (4,0) -- (4,4);
\node at (4.5,2) {$M$};
\draw[thick] (2,0) .. controls (2.5,2) .. (2,4);
\draw[thick] (2,0) .. controls ++(150:2) and ++(-70:1) .. (1.6,2) .. controls ++ (110:1) and ++ (-110:1) .. (2,4);
\node at (2.7,2) {$z$};
\node at (1,2) {$\Phi_\varepsilon(z)$};
\node at (2,1.5) {$\Sigma_\varepsilon$};
\draw [->] (2.12,0.5) -- +(200:0.3) node [anchor = south east] {$X$};
\draw [->] (2.23,1) -- +(170:0.35);
\draw [->] (2.35,2) -- +(190:0.3);
\draw [->] (2.32,2.5) -- +(180:0.5);
\draw [->] (2.23,3) -- +(170:0.3);
\draw [->] (2.13,3.5) -- +(150:0.2);
\end{tikzpicture} \caption{Variation of the path $z$ along a flow $\Phi_s$ whose tangent vector field on $z$ at $s=0$ coincides with $X$. The trajectories $\Phi_s(z)$ sweep the ribbon $\Sigma_\varepsilon$ as $s$ varies.}
        \label{fig:derivative_line}
    \end{figure}
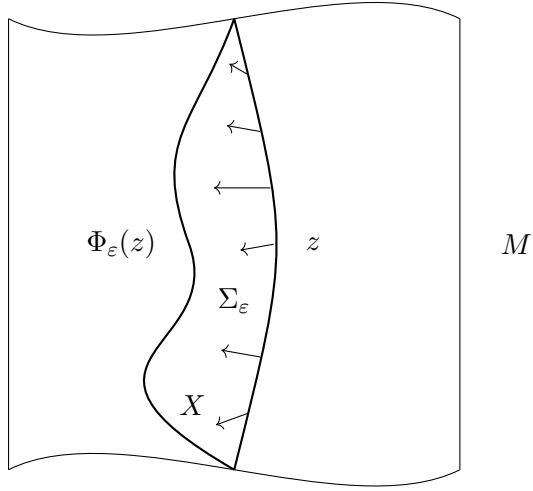
    
    At first sight, the scenario where the detector gap vanishes may look trivial from the point of view of the standard UDW detector. Indeed, in this case, the quantum Hamiltonian for the detector would simply be given by the interaction with the quantum field
    \begin{equation}
        H = \lambda \chi(\tau)\hat{\phi}(z(\tau))\sigma_x.
    \end{equation}
    The detector would be trivial in the sense that there would be no transition probability between the eigenstates of $\sigma_x$ for they would only evolve with a global phase. However, this interaction still evolves the eigenstates of $\sigma_z$, $\ket{g}$ and $\ket{e}$, in a non-trivial way. Measuring in this basis then allows one to observe the interference between the eigenstates of $\sigma_x$. Nevertheless, in order to talk about $\ket{g}$ and $\ket{e}$ for the detector one must be able to switch the energy gap on and off. One way of assigning meaning to the measurement of the $\sigma_z$ states in this model is to have a detector that starts with a non-zero energy gap $\Omega$ before interacting with the quantum field. This gap could be generated by an external field, for example. The detector could then be initialized in the ground state through a measurement. After that the gap is switched off during the interaction with the quantum field. Finally, the gap should be switched back on so that the transition from ground to excited state could be observed. 
    
    
    \section{Energy gap via a ``Twist''}\label{sec:massive}
    
    We have established a relationship between the gauge UDW detector and the Wilson lines of the field in terms of the defect \eqref{eq:derivative_defect_non_abelian} obtained at $\Omega=0$. To finish with our analysis, we would like to make this relationship explicit in the case of $\Omega\neq 0$. For this, we must revisit the analysis in section \ref{secLine}. Now we will restrain from including the energy gap
    \begin{equation}
        \int\dd{\tau}\frac{\Omega}{2}(\bar{\theta}\theta-\bar{\eta}\eta)
    \end{equation}
    into the propagator and instead treat it as an interaction term. In this scheme we can retain the gapless propagator at the cost of adding new interaction vertices
    \begin{equation}\label{eq:energy_vertices}
    \begin{aligned}
\begin{tikzpicture}[baseline]
\begin{feynman}
\vertex (in) at (-2,0);
\vertex[label = -90:\(\tau\), style = dot] (int) at (0,0) {};
\vertex (out) at (2,0);
\diagram*{
(in) -- [fermion] (int) -- [fermion] (out)
};
\end{feynman}
\end{tikzpicture}&=-\frac{\Omega(\tau)}{2},\\
\begin{tikzpicture}[baseline]
\begin{feynman}
\vertex (in) at (-2,0);
\vertex[label = -90:\(\tau\), style = dot] (int) at (0,0) {};
\vertex (out) at (2,0);
\diagram*{
(in) -- [charged scalar] (int) -- [charged scalar] (out)
};
\end{feynman}
\end{tikzpicture}&=\frac{\Omega(\tau)}{2}.
\end{aligned}
\end{equation}
In order to retain the reparametrization invariance of the model we must take $\Omega\dd{\tau}$ to be a 1-form, as we have discussed in Section \ref{secUDW}.  
Notice that there is a natural class of parametrizations where $\Omega$ is constant. For example, in the case where the energy gap of the detector in its own reference frame is a constant, one expects proper time to be one of these parametrizations. In particular, this argument can be generalized to define an analogous notion of proper time in theories without a spacetime metric.

A generic diagram can be constructed from building blocks of the form \eqref{eq:building_block}, where we now lift the restrictions on the external legs. We then obtain four types of building blocks corresponding to all possible combinations of these
\begin{equation}
    \begin{tikzpicture}[baseline]
    \begin{feynman}
    \vertex (in) at (-2,0) {\(a\)};
    \vertex[blob] (int) at (0,0) {};
    \vertex (out) at (2,0) {\(b\)};
    \diagram*{
    (in) -- [fermion] (int) -- [fermion] (out)
    };
    \end{feynman}
    \end{tikzpicture}, \quad \begin{tikzpicture}[baseline]
    \begin{feynman}
    \vertex (in) at (-2,0) {\(a\)};
    \vertex[blob] (int) at (0,0) {};
    \vertex (out) at (2,0) {\(b\)};
    \diagram*{
    (in) -- [charged scalar] (int) -- [charged scalar] (out)
    };
    \end{feynman}
    \end{tikzpicture},\label{eq:blocks_even}
\end{equation}
\begin{equation}
    \begin{tikzpicture}[baseline]
    \begin{feynman}
    \vertex (in) at (-2,0) {\(a\)};
    \vertex[blob] (int) at (0,0) {};
    \vertex (out) at (2,0) {\(b\)};
    \diagram*{
    (in) -- [fermion] (int) -- [charged scalar] (out)
    };
    \end{feynman}
    \end{tikzpicture}, \quad \begin{tikzpicture}[baseline]
    \begin{feynman}
    \vertex (in) at (-2,0) {\(a\)};
    \vertex[blob] (int) at (0,0) {};
    \vertex (out) at (2,0) {\(b\)};
    \diagram*{
    (in) -- [charged scalar] (int) -- [fermion] (out)
    };
    \end{feynman}
    \end{tikzpicture}.\label{eq:blocks_odd}
\end{equation}
Within these building blocks we will restrict the interactions to those of the detector with the field. These building blocks can then be attached to each other using the energy gap vertices found in \eqref{eq:energy_vertices} in such a way that the overall diagram takes the form \eqref{eq:building_block}, where now we do demand that the incoming leg is dashed while the outgoing leg is solid. Summing over all the possible diagrams constructed in this fashion will then yield the line defect for the gapped theory. 

Although we already know that the total result has the form \eqref{eq:UDW_defect} under the replacement $\phi(z)\mapsto \dot{z}^\mu \Tr_R(F_{\mu\nu}(z))X^\nu$, there is an alternative way of performing the sum of the diagrams described above. While this method only works in perturbation theory, it makes the connection with the Wilson lines of the theory explicit. First, we build one such diagram by choosing a list of building blocks out of the four available and joining them using our energy gap vertices. In this way, each building block will be bounded by either the initial point $\tau_i$, the end point $\tau_f$, or an energy gap vertex with some label $\lambda$. Due to the Heaviside functions in the propagators \eqref{eq:propagator}, the integrals in the interactions within a building block can be taken to be within the interval \textcolor{black}{bounding} it. As an example, consider the diagram
\begin{equation}\label{eq:example}
    \begin{tikzpicture}[baseline, scale = 0.6]
    \begin{feynman}
    \vertex (in) at (-6,0) {\(\tau_i\)};
    \vertex[label = -90:\(\tau_1^1\)] (int1) at (-4,0);
    \vertex (phi1) at (-4,1);
    \vertex[label = -90:\(\tau_2^1\)] (int2) at (-2,0);
    \vertex (phi2) at (-2,1);
    \vertex[label = -90:\(\lambda_1\), style = dot] (en1) at (0,0) {};
    \vertex[label = -90:\(\tau_1^2\)] (int3) at (2,0);
    \vertex (phi3) at (2,1);
    \vertex[label = -90:\(\tau_2^2\)] (int4) at (4,0);
    \vertex (phi4) at (4,1);
    \vertex[label = -90:\(\tau_3^2\)] (int5) at (6,0);
    \vertex (phi5) at (6,1);
    \vertex[label = -90:\(\tau_4^2\)] (int6) at (8,0);
    \vertex (phi6) at (8,1);
    \vertex[label = -90:\(\lambda_2\), style = dot] (en2) at (10,0) {};
    \vertex[label = -90:\(\tau_1^3\)] (int7) at (12,0);
    \vertex (phi7) at (12,1);
    \vertex[label = -90:\(\tau_2^3\)] (int8) at (14,0);
    \vertex (phi8) at (14,1);
    \vertex[label = -90:\(\tau_3^3\)] (int9) at (16,0);
    \vertex (phi9) at (16,1);
    \vertex (out) at (18,0) {\(\tau_f\)};
    \diagram*{
    (in) -- [charged scalar] (int1) -- [fermion] (int2) -- [charged scalar] (en1) -- [charged scalar] (int3) -- [fermion] (int4) -- [charged scalar] (int5) -- [fermion] (int6) -- [charged scalar] (en2) -- [charged scalar] (int7) -- [fermion] (int8) -- [charged scalar] (int9) -- [fermion] (out),
    (phi1) -- [gluon] (int1),
    (phi2) -- [gluon] (int2),
    (phi3) -- [gluon] (int3),
    (phi4) -- [gluon] (int4),
    (phi5) -- [gluon] (int5),
    (phi6) -- [gluon] (int6),
    (phi7) -- [gluon] (int7),
    (phi8) -- [gluon] (int8),
    (phi9) -- [gluon] (int9),
    };
    \end{feynman}
    \end{tikzpicture}.
\end{equation}
This diagram consists of three blocks joined by two energy gap vertices. The integrals appearing in the evaluation of the diagram
\begin{equation}
    \int\limits_{\tau_i}^{\tau_f}\dd{\tau_1^1}\int\limits_{\tau_1^1}^{\tau_f}\dd{\tau_2^1}\int\limits_{\tau_2^1}^{\tau_f}\dd{\lambda_1}\int\limits_{\lambda_1}^{\tau_f}\dd{\tau_1^2}\int\limits_{\tau_1^2}^{\tau_f}\dd{\tau_2^2}\int\limits_{\tau_2^2}^{\tau_f}\dd{\tau_3^2}\int\limits_{\tau_3^2}^{\tau_f}\dd{\tau_4^2}\int\limits_{\tau_4^2}^{\tau_f}\dd{\lambda_2}\int\limits_{\lambda_2}^{\tau_f}\dd{\tau_1^3}\int\limits_{\tau_1^3}^{\tau_f}\dd{\tau_2^3}\int\limits_{\tau_2^3}^{\tau_f}\dd{\tau_3^3},
\end{equation}
can be reordered into the form
\begin{equation}
    \int\limits_{\tau_i}^{\tau_f}\dd{\lambda_1}\int\limits_{\lambda_1}^{\tau_f}\dd{\lambda_2}\qty(\int\limits_{\tau_i}^{\lambda_1}\dd{\tau_1^1}\int\limits_{\tau_1^1}^{\lambda_1}\dd{\tau_2^1})\qty(\int\limits_{\lambda_1}^{\lambda_2}\dd{\tau_1^2}\int\limits_{\tau_1^2}^{\lambda_2}\dd{\tau_2^2}\int\limits_{\tau_2^2}^{\lambda_2}\dd{\tau_3^2}\int\limits_{\tau_3^2}^{\lambda_2}\dd{\tau_4^2})\qty(\int\limits_{\lambda_2}^{\tau_f}\dd{\tau_1^3}\int\limits_{\tau_1^3}^{\tau_f}\dd{\tau_2^3}\int\limits_{\tau_2^3}^{\tau_f}\dd{\tau_3^3}).
\end{equation}

Now that we have ordered the integrals in this fashion, we can sum over all possible interactions within each elementary block before performing the integrations over the energy gap vertex labels. In this way, each elementary block will take the form of a line defect. If the external legs of the block differ \eqref{eq:blocks_odd}, this line defect will take the form \eqref{eq:derivative_defect_non_abelian}, where the path $z$ has to be restricted to $[a,b]$, where $a$ and $b$ are the end points of the block in consideration. We will denote this defect $D^s_{a\rightarrow b}(A)$. On the other hand, if the external legs of the block coincide \eqref{eq:blocks_even}, the $\sinh$ in \eqref{eq:derivative_defect_non_abelian} is replaced by a $\cosh$. We will denote this defect by $D^c_{a\rightarrow b}(A)$. In our current example \eqref{eq:example} this summation will lead to the diagram
\begin{equation}\label{eq:example_2}
\begin{aligned}
    &\hphantom{{}={}}\begin{tikzpicture}[baseline]
    \begin{feynman}
    \vertex (in) at (-6,0) {\(\tau_i\)};
    \vertex[blob] (block1) at (-4,0) {};
    \vertex[label = -90:\(\lambda_1\), style = dot] (en1) at (-2,0) {};
    \vertex[blob] (block2) at (0,0) {};
    \vertex[label = -90:\(\lambda_2\), style = dot] (en2) at (2,0) {};
    \vertex[blob] (block3) at (4,0) {};
    \vertex (out) at (6,0) {\(\tau_f\)};
    \diagram*{
    (in) -- [charged scalar] (block1) -- [charged scalar] (en1) -- [charged scalar] (block2) -- [charged scalar] (en2) -- [charged scalar] (block3) -- [fermion] (out)};
    \end{feynman}
    \end{tikzpicture}\\
    &=\int_{\tau_i}^{\tau_f}\dd{\lambda_1}\frac{\Omega(\lambda_1)}{2}\int_{\lambda_1}^{\tau_f}\dd{\lambda_2}\frac{\Omega(\lambda_2)}{2} D^s_{\lambda_2\rightarrow \tau_f}(A)
    D^c_{\lambda_1\rightarrow\lambda_2}(A)D^c_{\tau_i\rightarrow\lambda_1}(A).
\end{aligned}
\end{equation}
In this way, we have obtained a method for computing the defect of the gapped gauge UDW detector in terms of \eqref{eq:derivative_defect_non_abelian} and its even analogue. It is obtained by sewing together the gapless defects using energy gap vertices.

\color{black}
\section{Conclusion}\label{secConc}

    We have presented a path integral formulation for the pointlike UDW model. The two-level system is described on the classical level by two Grassmann variables $\eta$ and $\theta$ and their respective conjugates. This ensures that the anti-commutation relations of the quantum system are preserved. With our formulation one can compute the transition probabilities according to sums over Feynman diagrams. This formulation is also fully covariant and naturally adapted to curved spacetimes, reinforcing the results of \cite{us} that the pointlike version of the UDW model is fully covariant.
    
    We have shown that the probability amplitude associated with UDW detectors can be related to a line defect. That is, we have provided a non-local effective observable of the quantum field supported along the detector's worldline which gives the excitation probability due to the interaction of the detector with the field. In the case of gauge theories this relationship is deeper, allowing the detector's transition probability to be associated with Wilson lines. In particular, we have explored a light-matter interaction inspired model and a non-abelian extension thereof, where we have shown that the line defect is given by the derivative of a Wilson line.
    
    Our work provides an extra toolbox for the study of particle detector models. Future applications of this could be related to the non-perturbative understanding of the associated defect that \eqref{eq:line_defect} yields.
    Moreover, this formulation opens the doors for particle detectors to be related to a variety of observables of quantum field theories in different scenarios. This is in tune with various recent proposals for physical setups exploring quantum-mechanical systems as probes for quantum fields. Examples include localized lasers which couple to Bose-Einstein condensates~\cite{UnruhBEC} and superconducting circuits coupled to a transmission line~\cite{EduardoSCC}, both of which turn out to be effectively described by the UDW model. The connection with line defects and Wilson loops also motivates further interest in probing gauge theories via particle detector models. One interesting question to pursue would be whether the confinement-deconfinement phase transition in Yang-Mills theories, for which Wilson loops provide an important order parameter~\cite{Wilson1974, tHooftPhaseTransition}, can be reinterpreted in terms of statements about local probes playing the role of detectors.

\section{Acknowledgements}

The authors would like to thank the PSI program for facilitating this research \textcolor{black}{and Kevin Costello for insightful discussions}. Research at Perimeter Institute is supported in part by the Government of Canada through the Department of Innovation, Science and Industry Canada and by the Province of Ontario through the Ministry of Colleges and Universities. T.R.P. thanks Drs. David Kubiznak and  Eduardo Martin-Martinez’s funding through their NSERC Discovery grants. B.S.L.T. acknowledges support from the Mike and Ophelia Lazaridis Fellowship.

\appendix

\section{Alternative derivation of the derivative of a Wilson line}\label{Bruno}

One can also derive the result in Eq.~\eqref{eq:linedefectderivative} as follows. In index-free notation, one can write

\begin{equation}\label{eq:A1}
    \int_{\tau_i}^{\tau_f} \dd\tau\,\dot{z}^\mu F_{\mu\nu}X^\nu = -\int_z i_X F,
\end{equation}
where the interior product $i_V\omega$ between a vector field $V$ and a $p$-form $\omega$ is defined in components as

\begin{equation}
    (i_V\omega)_{\nu_{1}\cdots\nu_{p-1}} = V^\mu \omega_{\mu\nu_{1}\cdots\nu_{p-1}}.
\end{equation}
We proceed by using Cartan's magic formula $\mathcal{L}_V\omega = \dd(i_V\omega) + i_V\dd\omega$ where $\mathcal{L}_V\omega$ corresponds to the Lie derivative of the $p$-form $\omega$, and is defined as the variation of $\omega$ along the flow generated by the vector field $V$. We then write $F = \dd A$ so that equation \eqref{eq:A1} can then be recast as
\begin{equation}\label{cartanmagicstep}
    \int_{\tau_i}^{\tau_f} \dd\tau\,\dot{z}^\mu F_{\mu\nu}X^\nu = \int_z \dd(i_X A) - \int_z \mathcal{L}_X A.
\end{equation}
Notice that, in principle, the vector field $X(\tau)$ is only defined along the curve $z(\tau)$ and not throughout spacetime. However, we can make sense of the manipulations above by considering a local extension of $X$. The first integral on the right-hand side of~\eqref{cartanmagicstep} reduces to a boundary term evaluated at the endpoints of the line. This can be set to zero by assuming that the coupling of the detector to $F_{\mu\nu}$ vanishes in those endpoints---for instance, if $X^\mu$ is taken to vanish before and after the interaction is turned on. From the definition of the Lie derivative, the remaining term can then be written as

\begin{equation}\label{eq:grownupboy}
    - \int_z \mathcal{L}_X A = - \int_z \dv{s}\Big|_{s=0}\Phi_s^*A=- \dv{s}\Big|_{s=0}\int_{\Phi_s(z)} A,
\end{equation}
where $s$ is the parameter along the flow $\Phi_s$ generated by the vector field $X$. Thus, even though the Lie derivative of $A$ with respect to $X$ in principle depends on the values of $X$ in a neighbourhood of the curve, the assumption that the vector field vanishes at the boundary allows us to conclude that \eqref{eq:grownupboy} is independent of the chosen extension. The expression of the line defect can then be recast as
 \begin{equation}
        D_z(A) = \sinh\left(-\dfrac{\dd}{\dd s}\Big|_{s=0} \log W_{\Phi_s(z)}(A)\right) = \sinh\left(-X\log W_{z}(A)\right),
    \end{equation}
as had been stated in Eq. \eqref{eq:derivative_defect}.

It is worth pointing out that an analogous calculation can also be performed in the case of non-Abelian gauge fields. We will schematically review the main steps below; the reader who is interested in following the calculation in more detail can refer to~\cite{Cherednikov2014, Polyakov1987}. In the non-Abelian case, the Wilson line is given by

\begin{equation}
    W_z(A) =  \mathcal{P}\exp\left(\int_z A\right),
\end{equation}
where $\mathcal{P}\exp$ corresponds to the path-ordered exponential. Its explicit expression can be put in the form

\begin{equation}
    W_z(A) =  \sum_{n=0}^\infty \int_{\tau_i}^{\tau_f}\dd \tau_1 \dots \int_{\tau_{n-2}}^{\tau_{n-1}}\dd\tau_n \left[\dot{z}^{\mu}(\tau_1)A_\mu(\tau_1)\right]\cdots\left[\dot{z}^{\nu}(\tau_n)A_\nu(\tau_n)\right].
\end{equation}
Computing $\dfrac{\dd}{\dd s}\Big|_{s=0}$ term by term in the series, we obtain via the Leibniz rule
\begin{equation}\label{wilsonlineexpansion}
    \dfrac{\dd}{\dd s}\Big|_{s=0}W_{\Phi_s(z)}(A) =  \sum_{n=0}^\infty \sum_{j=1}^n \int_{\tau_i}^{\tau_f}\dd \tau_1 \dots \int_{\tau_{n-2}}^{\tau_{n-1}}\dd\tau_n\,z^\ast A(\tau_1)\cdots z^\ast\left(\mathcal{L}_X A\right)(\tau_j)\cdots z^\ast A(\tau_n)
    \end{equation}
where we abbreviated $\dot{z}^{\mu}(\tau_i)V_\mu(\tau_i) = z^\ast V(\tau_i)$ for both the vector fields $A$ (for $i\neq j$) and $\mathcal{L}_X A$ (for $i=j$) in the expansion. By using Cartan's magic formula and rearranging the terms in the series, this can be put in the form
\begin{align}
    \dfrac{\dd}{\dd s}\Big|_{s=0}W_{\Phi_s}(A) = - \sum_{n=0}^\infty \sum_{j=1}^n \int_{\tau_i}^{\tau_f}\dd \tau_1 \dots \int_{\tau_{n-2}}^{\tau_{n-1}}\dd\tau_n\,& z^\ast A(\tau_1)\cdots z^\ast A(\tau_{j-1}) \nonumber\\
    \times&  \dot{z}^\mu(\tau_j)F_{\mu\nu}(\tau_j)X^\nu \nonumber \\
    \times&  z^\ast A(\tau_{j+1})\cdots z^\ast A(\tau_n),
\end{align}
where now $F_{\mu\nu} \equiv \partial_\mu A_\nu - \partial_\nu A_\mu - \comm{A_{\mu}}{A_\nu}$ is the non-Abelian field strength. The factors before and after the time when the field strength is being evaluated can then be further reorganized to give~\eqref{eq:non_abelian_derivative}, as stated in the main text.

\bibliography{references}

\begin{thebibliography}{45}
\expandafter\ifx\csname natexlab\endcsname\relax\def\natexlab#1{#1}\fi
\expandafter\ifx\csname bibnamefont\endcsname\relax
  \def\bibnamefont#1{#1}\fi
\expandafter\ifx\csname bibfnamefont\endcsname\relax
  \def\bibfnamefont#1{#1}\fi
\expandafter\ifx\csname citenamefont\endcsname\relax
  \def\citenamefont#1{#1}\fi
\expandafter\ifx\csname url\endcsname\relax
  \def\url#1{\texttt{#1}}\fi
\expandafter\ifx\csname urlprefix\endcsname\relax\def\urlprefix{URL }\fi
\providecommand{\bibinfo}[2]{#2}
\providecommand{\eprint}[2][]{\url{#2}}

\bibitem[{\citenamefont{Sorkin}(1956)}]{sorkin_1956}
\bibinfo{author}{\bibfnamefont{R.}~\bibnamefont{Sorkin}},
  \emph{\bibinfo{title}{Impossible Measurements on Quantum Fields}}
  (\bibinfo{publisher}{Cambridge University Press}, \bibinfo{year}{1956}),
  vol.~\bibinfo{volume}{2}, p. \bibinfo{pages}{293–305}.

\bibitem[{\citenamefont{Fewster and Verch}(2020)}]{fewster1}
\bibinfo{author}{\bibfnamefont{C.~J.} \bibnamefont{Fewster}} \bibnamefont{and}
  \bibinfo{author}{\bibfnamefont{R.}~\bibnamefont{Verch}},
  \bibinfo{journal}{Commun. Math. Phys.} \textbf{\bibinfo{volume}{378}},
  \bibinfo{pages}{851} (\bibinfo{year}{2020}),
  \urlprefix\url{https://doi.org/10.1007/s00220-020-03800-6}.

\bibitem[{\citenamefont{Fewster}(2020)}]{fewster2}
\bibinfo{author}{\bibfnamefont{C.~J.} \bibnamefont{Fewster}}, in
  \emph{\bibinfo{booktitle}{Progress and Visions in Quantum Theory in View of
  Gravity}}, edited by
  \bibinfo{editor}{\bibfnamefont{F.}~\bibnamefont{Finster}},
  \bibinfo{editor}{\bibfnamefont{D.}~\bibnamefont{Giulini}},
  \bibinfo{editor}{\bibfnamefont{J.}~\bibnamefont{Kleiner}}, \bibnamefont{and}
  \bibinfo{editor}{\bibfnamefont{J.}~\bibnamefont{Tolksdorf}}
  (\bibinfo{publisher}{Birkhäuser, Cham}, \bibinfo{year}{2020}), pp.
  \bibinfo{pages}{253--268},
  \urlprefix\url{https://doi.org/10.1007/978-3-030-38941-3_11}.

\bibitem[{\citenamefont{Bostelmann et~al.}(2020)\citenamefont{Bostelmann,
  Fewster, and Ruep}}]{fewster3}
\bibinfo{author}{\bibfnamefont{H.}~\bibnamefont{Bostelmann}},
  \bibinfo{author}{\bibfnamefont{C.~J.} \bibnamefont{Fewster}},
  \bibnamefont{and} \bibinfo{author}{\bibfnamefont{M.~H.} \bibnamefont{Ruep}},
  \emph{\bibinfo{title}{Impossible measurements require impossible apparatus}}
  (\bibinfo{year}{2020}), \eprint{2003.04660}.

\bibitem[{\citenamefont{Tong and Wong}(2014)}]{Tong2014}
\bibinfo{author}{\bibfnamefont{D.}~\bibnamefont{Tong}} \bibnamefont{and}
  \bibinfo{author}{\bibfnamefont{K.}~\bibnamefont{Wong}}, \bibinfo{journal}{J.
  High Energy Phys.} \textbf{\bibinfo{volume}{2014}}, \bibinfo{pages}{48}
  (\bibinfo{year}{2014}), ISSN \bibinfo{issn}{1029-8479},
  \urlprefix\url{https://doi.org/10.1007/JHEP06(2014)048}.

\bibitem[{\citenamefont{Takagi}(1986)}]{Takagi}
\bibinfo{author}{\bibfnamefont{S.}~\bibnamefont{Takagi}},
  \bibinfo{journal}{Prog. Theor. Phys. Supp.} \textbf{\bibinfo{volume}{88}},
  \bibinfo{pages}{1} (\bibinfo{year}{1986}), ISSN \bibinfo{issn}{0375-9687},
  \urlprefix\url{https://doi.org/10.1143/PTP.88.1}.

\bibitem[{\citenamefont{Schlicht}(2004)}]{Schlicht}
\bibinfo{author}{\bibfnamefont{S.}~\bibnamefont{Schlicht}},
  \bibinfo{journal}{Class. Quantum Gravity} \textbf{\bibinfo{volume}{21}},
  \bibinfo{pages}{4647} (\bibinfo{year}{2004}),
  \urlprefix\url{https://doi.org/10.1088%2F0264-9381%2F21%2F19%2F011}.

\bibitem[{\citenamefont{Ben-Benjamin et~al.}(2019)\citenamefont{Ben-Benjamin,
  Scully, Fulling, Lee, Page, Svidzinsky, Zubairy, Duff, Glauber, Schleich
  et~al.}}]{ScullyPage}
\bibinfo{author}{\bibfnamefont{J.~S.} \bibnamefont{Ben-Benjamin}},
  \bibinfo{author}{\bibfnamefont{M.~O.} \bibnamefont{Scully}},
  \bibinfo{author}{\bibfnamefont{S.~A.} \bibnamefont{Fulling}},
  \bibinfo{author}{\bibfnamefont{D.~M.} \bibnamefont{Lee}},
  \bibinfo{author}{\bibfnamefont{D.~N.} \bibnamefont{Page}},
  \bibinfo{author}{\bibfnamefont{A.~A.} \bibnamefont{Svidzinsky}},
  \bibinfo{author}{\bibfnamefont{M.~S.} \bibnamefont{Zubairy}},
  \bibinfo{author}{\bibfnamefont{M.~J.} \bibnamefont{Duff}},
  \bibinfo{author}{\bibfnamefont{R.}~\bibnamefont{Glauber}},
  \bibinfo{author}{\bibfnamefont{W.~P.} \bibnamefont{Schleich}},
  \bibnamefont{et~al.}, \bibinfo{journal}{Intl. J. Mod. Phys. A}
  \textbf{\bibinfo{volume}{34}}, \bibinfo{pages}{1941005}
  (\bibinfo{year}{2019}).

\bibitem[{\citenamefont{Unruh}(1976)}]{Unruh1976}
\bibinfo{author}{\bibfnamefont{W.~G.} \bibnamefont{Unruh}},
  \bibinfo{journal}{Phys. Rev. D} \textbf{\bibinfo{volume}{14}},
  \bibinfo{pages}{870} (\bibinfo{year}{1976}),
  \urlprefix\url{https://link.aps.org/doi/10.1103/PhysRevD.14.870}.

\bibitem[{\citenamefont{Hodgkinson et~al.}(2014)\citenamefont{Hodgkinson,
  Louko, and Ottewill}}]{Louko}
\bibinfo{author}{\bibfnamefont{L.}~\bibnamefont{Hodgkinson}},
  \bibinfo{author}{\bibfnamefont{J.}~\bibnamefont{Louko}}, \bibnamefont{and}
  \bibinfo{author}{\bibfnamefont{A.~C.} \bibnamefont{Ottewill}},
  \bibinfo{journal}{Phys. Rev. D} \textbf{\bibinfo{volume}{89}},
  \bibinfo{pages}{104002} (\bibinfo{year}{2014}),
  \urlprefix\url{https://link.aps.org/doi/10.1103/PhysRevD.89.104002}.

\bibitem[{\citenamefont{Wald}(1994)}]{Wald2}
\bibinfo{author}{\bibfnamefont{R.~M.} \bibnamefont{Wald}},
  \emph{\bibinfo{title}{Quantum Field Theory in Curved Spacetime and Black Hole
  Thermodynamics}} (\bibinfo{publisher}{The University of Chicago Press},
  \bibinfo{year}{1994}).

\bibitem[{\citenamefont{Unruh and Wald}(1984)}]{Unruh-Wald}
\bibinfo{author}{\bibfnamefont{W.~G.} \bibnamefont{Unruh}} \bibnamefont{and}
  \bibinfo{author}{\bibfnamefont{R.~M.} \bibnamefont{Wald}},
  \bibinfo{journal}{Phys. Rev. D} \textbf{\bibinfo{volume}{29}},
  \bibinfo{pages}{1047} (\bibinfo{year}{1984}).

\bibitem[{\citenamefont{Simidzija and Mart\'{\i}n-Mart\'{\i}nez}(2018)}]{Petar}
\bibinfo{author}{\bibfnamefont{P.}~\bibnamefont{Simidzija}} \bibnamefont{and}
  \bibinfo{author}{\bibfnamefont{E.}~\bibnamefont{Mart\'{\i}n-Mart\'{\i}nez}},
  \bibinfo{journal}{Phys. Rev. D} \textbf{\bibinfo{volume}{98}},
  \bibinfo{pages}{085007} (\bibinfo{year}{2018}),
  \urlprefix\url{https://link.aps.org/doi/10.1103/PhysRevD.98.085007}.

\bibitem[{\citenamefont{Pozas-Kerstjens and
  Mart\'{i}n-Mart\'{i}nez}(2015)}]{Pozas-Kerstjens:2015}
\bibinfo{author}{\bibfnamefont{A.}~\bibnamefont{Pozas-Kerstjens}}
  \bibnamefont{and}
  \bibinfo{author}{\bibfnamefont{E.}~\bibnamefont{Mart\'{i}n-Mart\'{i}nez}},
  \bibinfo{journal}{Phys. Rev. D} \textbf{\bibinfo{volume}{92}},
  \bibinfo{pages}{064042} (\bibinfo{year}{2015}),
  \urlprefix\url{http://link.aps.org/doi/10.1103/PhysRevD.92.064042}.

\bibitem[{\citenamefont{Pozas-Kerstjens and
  Mart\'{i}n-Mart\'{i}nez}(2016)}]{Pozas2016}
\bibinfo{author}{\bibfnamefont{A.}~\bibnamefont{Pozas-Kerstjens}}
  \bibnamefont{and}
  \bibinfo{author}{\bibfnamefont{E.}~\bibnamefont{Mart\'{i}n-Mart\'{i}nez}},
  \bibinfo{journal}{Phys. Rev. D} \textbf{\bibinfo{volume}{94}},
  \bibinfo{pages}{064074} (\bibinfo{year}{2016}),
  \urlprefix\url{https://link.aps.org/doi/10.1103/PhysRevD.94.064074}.

\bibitem[{\citenamefont{Henderson et~al.}(2018)\citenamefont{Henderson,
  Hennigar, Mann, Smith, and Zhang}}]{Henderson2018}
\bibinfo{author}{\bibfnamefont{L.~J.} \bibnamefont{Henderson}},
  \bibinfo{author}{\bibfnamefont{R.~A.} \bibnamefont{Hennigar}},
  \bibinfo{author}{\bibfnamefont{R.~B.} \bibnamefont{Mann}},
  \bibinfo{author}{\bibfnamefont{A.~R.~H.} \bibnamefont{Smith}},
  \bibnamefont{and} \bibinfo{author}{\bibfnamefont{J.}~\bibnamefont{Zhang}},
  \bibinfo{journal}{Class. Quantum Gravity} \textbf{\bibinfo{volume}{35}},
  \bibinfo{pages}{21LT02} (\bibinfo{year}{2018}),
  \urlprefix\url{https://doi.org/10.1088%2F1361-6382%2Faae27e}.

\bibitem[{\citenamefont{Trevison et~al.}(2018)\citenamefont{Trevison,
  Yamaguchi, and Hotta}}]{hottaHarvesting}
\bibinfo{author}{\bibfnamefont{J.}~\bibnamefont{Trevison}},
  \bibinfo{author}{\bibfnamefont{K.}~\bibnamefont{Yamaguchi}},
  \bibnamefont{and} \bibinfo{author}{\bibfnamefont{M.}~\bibnamefont{Hotta}},
  \bibinfo{journal}{PTEP} \textbf{\bibinfo{volume}{2018}},
  \bibinfo{pages}{103A03} (\bibinfo{year}{2018}), \eprint{1808.01764}.

\bibitem[{\citenamefont{Funai et~al.}(2019)\citenamefont{Funai, Louko, and
  Mart\'{i}n-Mart\'{i}nez}}]{Nicho1}
\bibinfo{author}{\bibfnamefont{N.}~\bibnamefont{Funai}},
  \bibinfo{author}{\bibfnamefont{J.}~\bibnamefont{Louko}}, \bibnamefont{and}
  \bibinfo{author}{\bibfnamefont{E.}~\bibnamefont{Mart\'{i}n-Mart\'{i}nez}},
  \bibinfo{journal}{Phys. Rev. D} \textbf{\bibinfo{volume}{99}},
  \bibinfo{pages}{065014} (\bibinfo{year}{2019}),
  \urlprefix\url{https://link.aps.org/doi/10.1103/PhysRevD.99.065014}.

\bibitem[{\citenamefont{Lopp and Martín-Martínez}(2020)}]{richard}
\bibinfo{author}{\bibfnamefont{R.}~\bibnamefont{Lopp}} \bibnamefont{and}
  \bibinfo{author}{\bibfnamefont{E.}~\bibnamefont{Martín-Martínez}},
  \emph{\bibinfo{title}{Quantum delocalization, gauge and quantum optics: The
  light-matter interaction in relativistic quantum information}}
  (\bibinfo{year}{2020}), \eprint{2008.12785}.

\bibitem[{\citenamefont{Faure et~al.}(2020)\citenamefont{Faure, Perche, and
  Torres}}]{remi}
\bibinfo{author}{\bibfnamefont{R.}~\bibnamefont{Faure}},
  \bibinfo{author}{\bibfnamefont{T.~R.} \bibnamefont{Perche}},
  \bibnamefont{and} \bibinfo{author}{\bibfnamefont{B.~d. S.~L.}
  \bibnamefont{Torres}}, \bibinfo{journal}{Phys. Rev. D}
  \textbf{\bibinfo{volume}{101}}, \bibinfo{pages}{125018}
  (\bibinfo{year}{2020}),
  \urlprefix\url{https://link.aps.org/doi/10.1103/PhysRevD.101.125018}.

\bibitem[{\citenamefont{Torres et~al.}(2020)\citenamefont{Torres, Perche,
  Landulfo, and Matsas}}]{neutrinoOscillations}
\bibinfo{author}{\bibfnamefont{B.~d. S.~L.} \bibnamefont{Torres}},
  \bibinfo{author}{\bibfnamefont{T.~R.} \bibnamefont{Perche}},
  \bibinfo{author}{\bibfnamefont{A.~G.~S.} \bibnamefont{Landulfo}},
  \bibnamefont{and} \bibinfo{author}{\bibfnamefont{G.~E.~A.}
  \bibnamefont{Matsas}}, \bibinfo{journal}{Phys. Rev. D}
  \textbf{\bibinfo{volume}{102}}, \bibinfo{pages}{093003}
  (\bibinfo{year}{2020}),
  \urlprefix\url{https://link.aps.org/doi/10.1103/PhysRevD.102.093003}.

\bibitem[{\citenamefont{DeWitt}(1980)}]{DeWitt}
\bibinfo{author}{\bibfnamefont{B.}~\bibnamefont{DeWitt}},
  \emph{\bibinfo{title}{General Relativity; an Einstein Centenary Survey}}
  (\bibinfo{publisher}{Cambridge University Press},
  \bibinfo{address}{Cambridge, UK}, \bibinfo{year}{1980}).

\bibitem[{\citenamefont{Mart\'{\i}n-Mart\'{\i}nez
  et~al.}(2020)\citenamefont{Mart\'{\i}n-Mart\'{\i}nez, Perche, and
  de~S.~L.~Torres}}]{us}
\bibinfo{author}{\bibfnamefont{E.}~\bibnamefont{Mart\'{\i}n-Mart\'{\i}nez}},
  \bibinfo{author}{\bibfnamefont{T.~R.} \bibnamefont{Perche}},
  \bibnamefont{and}
  \bibinfo{author}{\bibfnamefont{B.}~\bibnamefont{de~S.~L.~Torres}},
  \bibinfo{journal}{Phys. Rev. D} \textbf{\bibinfo{volume}{101}},
  \bibinfo{pages}{045017} (\bibinfo{year}{2020}),
  \urlprefix\url{https://link.aps.org/doi/10.1103/PhysRevD.101.045017}.

\bibitem[{\citenamefont{Kukita and Nambu}(2017)}]{Kukita2017}
\bibinfo{author}{\bibfnamefont{S.}~\bibnamefont{Kukita}} \bibnamefont{and}
  \bibinfo{author}{\bibfnamefont{Y.}~\bibnamefont{Nambu}},
  \bibinfo{journal}{Class. Quantum Gravity} \textbf{\bibinfo{volume}{34}},
  \bibinfo{pages}{235010} (\bibinfo{year}{2017}),
  \urlprefix\url{https://doi.org/10.1088/1361-6382/aa8e31}.

\bibitem[{\citenamefont{Henderson et~al.}(2019)\citenamefont{Henderson,
  Hennigar, Mann, Smith, and Zhang}}]{Henderson2019}
\bibinfo{author}{\bibfnamefont{L.~J.} \bibnamefont{Henderson}},
  \bibinfo{author}{\bibfnamefont{R.~A.} \bibnamefont{Hennigar}},
  \bibinfo{author}{\bibfnamefont{R.~B.} \bibnamefont{Mann}},
  \bibinfo{author}{\bibfnamefont{A.~R.~H.} \bibnamefont{Smith}},
  \bibnamefont{and} \bibinfo{author}{\bibfnamefont{J.}~\bibnamefont{Zhang}},
  \bibinfo{journal}{J. High Energy Phys.} \textbf{\bibinfo{volume}{2019}},
  \bibinfo{pages}{178} (\bibinfo{year}{2019}), ISSN \bibinfo{issn}{1029-8479},
  \urlprefix\url{https://doi.org/10.1007/JHEP05(2019)178}.

\bibitem[{\citenamefont{Ng et~al.}(2018{\natexlab{a}})\citenamefont{Ng, Mann,
  and Mart\'{\i}n-Mart\'{\i}nez}}]{Ng1}
\bibinfo{author}{\bibfnamefont{K.~K.} \bibnamefont{Ng}},
  \bibinfo{author}{\bibfnamefont{R.~B.} \bibnamefont{Mann}}, \bibnamefont{and}
  \bibinfo{author}{\bibfnamefont{E.}~\bibnamefont{Mart\'{\i}n-Mart\'{\i}nez}},
  \bibinfo{journal}{Phys. Rev. D} \textbf{\bibinfo{volume}{97}},
  \bibinfo{pages}{125011} (\bibinfo{year}{2018}{\natexlab{a}}),
  \urlprefix\url{https://link.aps.org/doi/10.1103/PhysRevD.97.125011}.

\bibitem[{\citenamefont{Ng et~al.}(2018{\natexlab{b}})\citenamefont{Ng, Mann,
  and Mart\'{\i}n-Mart\'{\i}nez}}]{Ng2}
\bibinfo{author}{\bibfnamefont{K.~K.} \bibnamefont{Ng}},
  \bibinfo{author}{\bibfnamefont{R.~B.} \bibnamefont{Mann}}, \bibnamefont{and}
  \bibinfo{author}{\bibfnamefont{E.}~\bibnamefont{Mart\'{\i}n-Mart\'{\i}nez}},
  \bibinfo{journal}{Phys. Rev. D} \textbf{\bibinfo{volume}{98}},
  \bibinfo{pages}{125005} (\bibinfo{year}{2018}{\natexlab{b}}),
  \urlprefix\url{https://link.aps.org/doi/10.1103/PhysRevD.98.125005}.

\bibitem[{\citenamefont{Martín-Martínez
  et~al.}(2020)\citenamefont{Martín-Martínez, Perche, and
  de~S.~L.~Torres}}]{us2}
\bibinfo{author}{\bibfnamefont{E.}~\bibnamefont{Martín-Martínez}},
  \bibinfo{author}{\bibfnamefont{T.~R.} \bibnamefont{Perche}},
  \bibnamefont{and}
  \bibinfo{author}{\bibfnamefont{B.}~\bibnamefont{de~S.~L.~Torres}},
  \emph{\bibinfo{title}{Broken covariance of particle detector models in
  relativistic quantum information}} (\bibinfo{year}{2020}),
  \eprint{2006.12514}.

\bibitem[{\citenamefont{Mart\'{i}n-Mart\'{i}nez and
  Rodriguez-Lopez}(2018)}]{eduardo}
\bibinfo{author}{\bibfnamefont{E.}~\bibnamefont{Mart\'{i}n-Mart\'{i}nez}}
  \bibnamefont{and}
  \bibinfo{author}{\bibfnamefont{P.}~\bibnamefont{Rodriguez-Lopez}},
  \bibinfo{journal}{Phys. Rev. D} \textbf{\bibinfo{volume}{97}},
  \bibinfo{pages}{105026} (\bibinfo{year}{2018}),
  \urlprefix\url{https://link.aps.org/doi/10.1103/PhysRevD.97.105026}.

\bibitem[{\citenamefont{Zinn-Justin}(2004)}]{Zinn-Justin2004}
\bibinfo{author}{\bibfnamefont{J.}~\bibnamefont{Zinn-Justin}},
  \emph{\bibinfo{title}{{Path Integrals in Quantum Mechanics}}}
  (\bibinfo{publisher}{Oxford University Press}, \bibinfo{year}{2004}), ISBN
  \bibinfo{isbn}{9780198566748},
  \urlprefix\url{https://oxford.universitypressscholarship.com/view/10.1093/acprof:oso/9780198566748.001.0001/acprof-9780198566748}.

\bibitem[{\citenamefont{Lawson and Michelson}(1989)}]{Lawson1989}
\bibinfo{author}{\bibfnamefont{H.~B.} \bibnamefont{Lawson}} \bibnamefont{and}
  \bibinfo{author}{\bibfnamefont{M.-L.} \bibnamefont{Michelson}},
  \emph{\bibinfo{title}{{Spin Geometry}}} (\bibinfo{publisher}{Princeton
  University Press}, \bibinfo{year}{1989}), ISBN \bibinfo{isbn}{9780691085425}.

\bibitem[{\citenamefont{Merad et~al.}(2001)\citenamefont{Merad, Boudjedaa, and
  Chetouani}}]{Merad2001}
\bibinfo{author}{\bibfnamefont{M.}~\bibnamefont{Merad}},
  \bibinfo{author}{\bibfnamefont{T.}~\bibnamefont{Boudjedaa}},
  \bibnamefont{and}
  \bibinfo{author}{\bibfnamefont{L.}~\bibnamefont{Chetouani}},
  \bibinfo{journal}{Journal of the Korean Physical Society}
  \textbf{\bibinfo{volume}{38}}, \bibinfo{pages}{69} (\bibinfo{year}{2001}),
  ISSN \bibinfo{issn}{03744884}.

\bibitem[{\citenamefont{Aouachria and Chetouani}(2002)}]{Aouachria2002}
\bibinfo{author}{\bibfnamefont{M.}~\bibnamefont{Aouachria}} \bibnamefont{and}
  \bibinfo{author}{\bibfnamefont{L.}~\bibnamefont{Chetouani}},
  \bibinfo{journal}{The European Physical Journal C}
  \textbf{\bibinfo{volume}{25}}, \bibinfo{pages}{333} (\bibinfo{year}{2002}),
  ISSN \bibinfo{issn}{1434-6044},
  \urlprefix\url{http://link.springer.com/10.1007/s10052-002-0984-0}.

\bibitem[{\citenamefont{Aouachria}(2009)}]{Aouachria2009}
\bibinfo{author}{\bibfnamefont{M.}~\bibnamefont{Aouachria}},
  \textbf{\bibinfo{volume}{2}}, \bibinfo{pages}{1} (\bibinfo{year}{2009}),
  \urlprefix\url{http://www.physicsegypt.org/nuppac09/npc9035.pdf}.

\bibitem[{\citenamefont{Aouachria}(2013)}]{Aouachria2013}
\bibinfo{author}{\bibfnamefont{M.}~\bibnamefont{Aouachria}},
  \bibinfo{journal}{Journal of Physics: Conference Series}
  \textbf{\bibinfo{volume}{435}} (\bibinfo{year}{2013}), ISSN
  \bibinfo{issn}{17426596}.

\bibitem[{\citenamefont{Gaiotto et~al.}(2015)\citenamefont{Gaiotto, Kapustin,
  Seiberg, and Willett}}]{queTrampoSpringer}
\bibinfo{author}{\bibfnamefont{D.}~\bibnamefont{Gaiotto}},
  \bibinfo{author}{\bibfnamefont{A.}~\bibnamefont{Kapustin}},
  \bibinfo{author}{\bibfnamefont{N.}~\bibnamefont{Seiberg}}, \bibnamefont{and}
  \bibinfo{author}{\bibfnamefont{B.}~\bibnamefont{Willett}},
  \bibinfo{journal}{J. High Energy Phys.} \textbf{\bibinfo{volume}{2015}},
  \bibinfo{pages}{172} (\bibinfo{year}{2015}), ISSN \bibinfo{issn}{1029-8479},
  \urlprefix\url{https://doi.org/10.1007/JHEP02(2015)172}.

\bibitem[{\citenamefont{Giles}(1981)}]{ivaan}
\bibinfo{author}{\bibfnamefont{R.}~\bibnamefont{Giles}},
  \bibinfo{journal}{Phys. Rev. D} \textbf{\bibinfo{volume}{24}},
  \bibinfo{pages}{2160} (\bibinfo{year}{1981}),
  \urlprefix\url{https://link.aps.org/doi/10.1103/PhysRevD.24.2160}.

\bibitem[{\citenamefont{Tong}(2018)}]{Tong2018}
\bibinfo{author}{\bibfnamefont{D.}~\bibnamefont{Tong}},
  \emph{\bibinfo{title}{{Gauge Theory}}} (\bibinfo{year}{2018}),
  \urlprefix\url{http://www.damtp.cam.ac.uk/user/tong/gaugetheory.html}.

\bibitem[{\citenamefont{Göppert-Mayer}(2009)}]{MariaGoeppert-Meyer}
\bibinfo{author}{\bibfnamefont{M.}~\bibnamefont{Göppert-Mayer}},
  \bibinfo{journal}{Annalen der Physik} \textbf{\bibinfo{volume}{18}},
  \bibinfo{pages}{466} (\bibinfo{year}{2009}),
  \eprint{https://onlinelibrary.wiley.com/doi/pdf/10.1002/andp.200910358},
  \urlprefix\url{https://onlinelibrary.wiley.com/doi/abs/10.1002/andp.200910358}.

\bibitem[{\citenamefont{Cherednikov et~al.}(2014)\citenamefont{Cherednikov,
  Mertens, and Van~der Veken}}]{Cherednikov2014}
\bibinfo{author}{\bibfnamefont{I.~O.} \bibnamefont{Cherednikov}},
  \bibinfo{author}{\bibfnamefont{T.}~\bibnamefont{Mertens}}, \bibnamefont{and}
  \bibinfo{author}{\bibfnamefont{F.~F.} \bibnamefont{Van~der Veken}},
  \emph{\bibinfo{title}{{Wilson Lines in Quantum Field Theory}}},
  \bibinfo{number}{2018} (\bibinfo{publisher}{De Gruyter},
  \bibinfo{year}{2014}), ISBN \bibinfo{isbn}{9783110309218},
  \urlprefix\url{https://www.degruyter.com/view/title/301063}.

\bibitem[{\citenamefont{Gooding et~al.}(2020)\citenamefont{Gooding, Biermann,
  Erne, Louko, Unruh, Schmiedmayer, and Weinfurtner}}]{UnruhBEC}
\bibinfo{author}{\bibfnamefont{C.}~\bibnamefont{Gooding}},
  \bibinfo{author}{\bibfnamefont{S.}~\bibnamefont{Biermann}},
  \bibinfo{author}{\bibfnamefont{S.}~\bibnamefont{Erne}},
  \bibinfo{author}{\bibfnamefont{J.}~\bibnamefont{Louko}},
  \bibinfo{author}{\bibfnamefont{W.~G.} \bibnamefont{Unruh}},
  \bibinfo{author}{\bibfnamefont{J.}~\bibnamefont{Schmiedmayer}},
  \bibnamefont{and}
  \bibinfo{author}{\bibfnamefont{S.}~\bibnamefont{Weinfurtner}},
  \bibinfo{journal}{Phys. Rev. Lett.} \textbf{\bibinfo{volume}{125}},
  \bibinfo{pages}{213603} (\bibinfo{year}{2020}),
  \urlprefix\url{https://link.aps.org/doi/10.1103/PhysRevLett.125.213603}.

\bibitem[{\citenamefont{McKay et~al.}(2017)\citenamefont{McKay, Lupascu, and
  Mart\'{\i}n-Mart\'{\i}nez}}]{EduardoSCC}
\bibinfo{author}{\bibfnamefont{E.}~\bibnamefont{McKay}},
  \bibinfo{author}{\bibfnamefont{A.}~\bibnamefont{Lupascu}}, \bibnamefont{and}
  \bibinfo{author}{\bibfnamefont{E.}~\bibnamefont{Mart\'{\i}n-Mart\'{\i}nez}},
  \bibinfo{journal}{Phys. Rev. A} \textbf{\bibinfo{volume}{96}},
  \bibinfo{pages}{052325} (\bibinfo{year}{2017}),
  \urlprefix\url{https://link.aps.org/doi/10.1103/PhysRevA.96.052325}.

\bibitem[{\citenamefont{Wilson}(1974)}]{Wilson1974}
\bibinfo{author}{\bibfnamefont{K.~G.} \bibnamefont{Wilson}},
  \bibinfo{journal}{Phys. Rev. D} \textbf{\bibinfo{volume}{10}},
  \bibinfo{pages}{2445} (\bibinfo{year}{1974}),
  \urlprefix\url{https://link.aps.org/doi/10.1103/PhysRevD.10.2445}.

\bibitem[{\citenamefont{{'t Hooft}}(1978)}]{tHooftPhaseTransition}
\bibinfo{author}{\bibfnamefont{G.}~\bibnamefont{{'t Hooft}}},
  \bibinfo{journal}{Nuclear Physics B} \textbf{\bibinfo{volume}{138}},
  \bibinfo{pages}{1 } (\bibinfo{year}{1978}), ISSN \bibinfo{issn}{0550-3213},
  \urlprefix\url{http://www.sciencedirect.com/science/article/pii/0550321378901530}.

\bibitem[{\citenamefont{Polyakov}(1987)}]{Polyakov1987}
\bibinfo{author}{\bibfnamefont{A.}~\bibnamefont{Polyakov}},
  \emph{\bibinfo{title}{{Gauge Fields and Strings}}}
  (\bibinfo{publisher}{Routledge}, \bibinfo{year}{1987}), ISBN
  \bibinfo{isbn}{9780203755082},
  \urlprefix\url{https://www.taylorfrancis.com/books/9781351446099}.

\end{thebibliography}

\end{document}